\documentclass[aps,showpacs,preprintnumbers,amsmath,amssymb]{revtex4}
\usepackage{dcolumn}
\usepackage[dvips]{epsfig}
\usepackage{graphics, epsfig}

\begin{document}
\baselineskip=0.8 cm
\title{\bf Energy extraction from a Konoplya-Zhidenko  rotating non-Kerr black hole}

\author{Fen Long$^{1}$,  Songbai Chen$^{1, 2, 3}$\footnote{Corresponding author: csb3752@hunnu.edu.cn}, Shangyun Wang $^{1}$, Jiliang
Jing$^{1, 2, 3}$ \footnote{jljing@hunnu.edu.cn}}

\affiliation{$^{\textit{1}}$Institute of Physics and Department of Physics,  Hunan
Normal University,   Changsha,  Hunan 410081,  People's Republic of
China \\ $^{\textit{2}}$Key Laboratory of Low Dimensional Quantum Structures \\
and Quantum Control of Ministry of Education,  Hunan Normal
University,  Changsha,  Hunan 410081,  People's Republic of China\\
$^{\textit{3}}$Synergetic Innovation Center for Quantum Effects and Applications,
Hunan Normal University,  Changsha,  Hunan 410081,  People's Republic
of China}

\begin{abstract}
\baselineskip=0.6 cm
\begin{center}
{\bf Abstract}
\end{center}

We have investigated the properties of the ergosphere and the energy extraction by Penrose process in a Konoplya-Zhidenko rotating non-Kerr black hole spacetime. We find that the ergosphere becomes thin and the maximum efficiency of energy extraction decreases as the deformation parameter increases. For the case with $a<M$, the positive deformation parameter yields that the maximum efficiency is lower than that in the Kerr black hole with the same rotation parameter. However, for the superspinning case with $a>M$, we find that the maximum efficiency can reach so high that it is almost unlimited as the positive deformation parameter is close to zero, which is a new feature of energy extraction in such kind of rotating non-Kerr black hole spacetime.

\end{abstract}
\maketitle

\newpage
\section{Introduction}

The existence of black hole in our Universe was confirmed by the recently reported gravitational wave events, such as GW150914, GW151226 and GW170104 \cite{W2,W21,W31}. In Einstein's theory of gravity, a neutral rotating astrophysical black hole in vacuum is described completely by the Kerr metric only with two parameters, the mass $M$ and the rotation parameter $a$, which is supported by the no-hair theorem \cite{L1}. For a Kerr black hole, there is
a fundamental limit (i.e., $a< M$ ) from the weak cosmic censorship conjecture, which guarantees that the central singularity is always covered by the event horizon. Although Einstein's General Relativity
passed a series of observational and experimental tests \cite{W1,W3}, it is worth noting that there is still ample room for other alternative theories of gravity since the current observations including the recent gravitational wave events \cite{W2,W21,W31} cannot completely exclude the possibility of the deviation from Einstein's gravity theory. Especially, in cosmology, modifying Einstein's theory of gravity \cite{fR} is still one of the most promising ways to explain the accelerating expansion of the current Universe observed through astronomical experiments \cite{sd1,sd2,sd3,sd4,sd5} since it does not resort to the exotic component, such as dark energy in General Relativity.
Therefore, it is still necessary to study the black hole solutions in other alternative theories of gravity.

The rotating non-Kerr spacetimes are described by a kind of deformed Kerr-like metrics, which can be treated as  vacuum solutions of a unknown alternative theory of gravity beyond Einstein¡¯s general relativity. Besides the mass $M$ and rotation parameter $a$, these spacetimes possess an extra deformation parameter describing the deviation
from the usual Kerr one. Choosing certain a proper deformation function, Johannsen and Psaltis \cite{JP1} constructed a rotating non-Kerr metric to  examine the no-hair theorem. Johannsen-Psaltis non-Kerr spacetime is one of the most important non-Kerr spacetimes with the same asymptotic behaviors of Kerr spacetime in the far-field region. However, the presence of the deformation parameter changes the structures of spacetime in the strong-field region. The horizon radius is a function of the polar angle $\theta$ and the horizon surface does not possess the spherical topology as the value of deformation parameter lies within a certain range \cite{JP1,CBa1}. Moreover, there is no restriction on
the value of the rotational parameter $a$ for the Johannsen-Psaltis non-Kerr spacetime. It means that this metric can describe a superspining black hole ( $a>M$ ) which is absence  in the Einstein's General Relativity.
The special observable effects in Johannsen-Psaltis non-Kerr spacetime are studied in \cite{Cos101,Cos102,chen1,chen12,Kraw,Ra1,Fa1,And1}, and the constraints to the possible deviation are also made by using of the observation data from the quasi-periodic oscillations \cite{Test3,Test302,Test303,Test304,Test305} and the continuum-fitting and iron line \cite{Test1,Test2,Test201,Test202}.

Another important rotating non-Kerr black hole metric is proposed by Konoplya and Zhidenko through adding a static deformation \cite{W4}. Their main purpose is to check whether the detection of gravitational waves can leave a window open for alternative theories. With this rotating non-Kerr metric, they found that some non-negligible deviation from the Kerr spacetime can also yield the same frequencies of the black-hole ringing. Moreover, the constraints from quasi-periodic oscillations \cite{GKt01} and the iron line \cite{GKt02} also support that
a real astrophysical black hole  could be described by Konoplya-Zhidenko rotating non-Kerr metric. Although a Konoplya-Zhidenko non-Kerr black hole metric has many similar properties of Johannsen-Psaltis non-Kerr one, but its horizon radius is independent of the polar angle $\theta$ and the horizon surface still remains spherical structure as in Kerr case \cite{W4}.
The study of strong gravitational lensing  shows that there are some distinct features differed from those in Kerr and Johannsen-Psaltis non-Kerr spacetimes \cite{W6}.

The Penrose process \cite{L10,L11,L12} is an important method to extract energy from a rotation black hole, which has been proposed to explain the formation of the power jets and the power energy for a active galactic nuclei, X-ray binaries and quasars.
With Penrose process, the energy extraction has been studied in the Johannsen-Psaltis non-Kerr spacetime \cite{chen2,chen201}, which indicates that the deformation parameter enhances the maximum efficiency of the energy extraction process greatly. In this paper, we will investigate in detail the ergosphere of the Konoplya-Zhidenko non-Kerr black
hole and probe the effects of the deformation parameter on the negative
energy state and the efficiency of the energy extraction in this non-Kerr spacetime. Moreover, we will explore how it differs from those in the
Johannsen-Psaltis rotating non-Kerr case.

The paper is organized as follows. In Sec. II, we review briefly the metric of the Konoplya-Zhidenko rotating non-Kerr black hole  and then analyze its ergosphere structure. In Sec. III, we investigate the efficiency of the energy extraction by using the Penrose process. We end the paper with a summary

\section{Horizons and infinite redshift surfaces of a Konoplya-Zhidenko rotating non-Kerr black hole}

In the Boyer-Lindquist coordinates, the metric of a Konoplya-Zhidenko rotating non-Kerr black hole can be expressed as \cite{W4}
\begin{eqnarray}
ds^2&=&-\frac{N^2(r, \theta)-W^2(r, \theta)\sin^2\theta}{K^2(r, \theta)}dt^2
-{2rW(r, \theta)\sin^2\theta}dtd\phi+{K^2(r, \theta)r^2\sin^2\theta}d\phi^2
\nonumber\\
&+&\Sigma(r, \theta)\bigg[{\frac{B^2(r, \theta)}{N^2(r, \theta)}+r^2d\theta^2}\bigg], \label{metric1}
\end{eqnarray}
with
\begin{eqnarray}
N^2(r, \theta)&=&\frac{r^2-2Mr+a^2}{r^2}-\frac{\eta}{r^3}, \;\;\;\;
B(r, \theta)=1, \;\;\;\;
\Sigma(r, \theta)=\frac{r^2+a^2\cos^2\theta}{r^2}, \nonumber\\
K^2(r, \theta)&=&\frac{(r^2+a^2)^2-a^2\sin^2\theta(r^2-2Mr+a^2)}{r^2(r^2+a^2\cos^2\theta)}
+\frac{a^2\eta\sin^2\theta}{r^3(r^2+a^2\cos^2\theta)}, \nonumber\\
W(r, \theta)&=&\frac{2Ma}{r^2+a^2\cos^2\theta}
+\frac{\eta a}{r^2(r^2+a^2\cos^2\theta)}.
\end{eqnarray}
Like a usual rotating non-Kerr metric, it can be obtained through deforming the Kerr metric.  Here $M$ is the mass of black hole and $a$ is the rotation parameter. The deformation parameter $\eta$ describes the deviations from the Kerr spacetime. As the deformation parameter disappears, this metric can be reduced to that of usual Kerr black hole. Obviously, the presence of
the deformation parameter $\eta$ does not change the asymptotic structure of spacetime at the spatial infinite, but modifies the behavior of spacetime  in the strong field region. However, due to the difference between deformation, the Konoplya-Zhidenko rotating non-Kerr black hole possesses some special properties of spacetime differed from those of Johannsen-Psaltis non-Kerr one \cite{W4,W6}.

The horizons of the black hole are defined by the equation
\begin{eqnarray}
r^3-2Mr^2+a^2r-\eta=0. \label{metric5}
\end{eqnarray}
which gives
\begin{eqnarray}
r^{1}_{H}&=&\frac{1}{3}\bigg[2M+\frac{2^{1/3}(4M^2- 3a^2)}{\mathcal{A}^{1/3}}+
\frac{\mathcal{A}^{1/3}}{2^{1/3}}\bigg],\label{h1}\\
r^2_{H}&=&\frac{1}{3}\bigg[2M-\frac{(1+\sqrt{3}i)(4M^2- 3a^2)}{2^{2/3}\mathcal{A}^{1/3}}-
\frac{(1-\sqrt{3}i)\mathcal{A}^{1/3}}{2^{4/3}}\bigg],\label{h2}\\
r^{3}_{H}&=&\frac{1}{3}\bigg[2M-\frac{(1-\sqrt{3}i)(4M^2- 3a^2)}{2^{2/3}\mathcal{A}^{1/3}}-
\frac{(1+\sqrt{3}i)\mathcal{A}^{1/3}}{2^{4/3}}\bigg],\label{h3}
\end{eqnarray}
with
\begin{eqnarray}
\mathcal{A}=16M^3+27\eta-18 a^2M+\sqrt{(16M^3+27\eta-18a^2M)^2-4(4M^2-3a^2)^3}.
\end{eqnarray}
Differing from a Johannsen-Psaltis non-Kerr case, the horizon radius of a Konoplya-Zhidenko rotating non-Kerr black hole is independent of the polar angle $\theta$ and its horizon surface still remains spherical structure as in Kerr case.
However, the presence of the deformation parameter $\eta$ changes the condition of the existence of black hole horizons and further affects the number and positions of horizons. Through analysing Eqs.(\ref{h1})-(\ref{h3}), one can find that there exist two threshold values for the existence of horizon in the spacetime \cite{W6}
\begin{eqnarray}\label{etam0}
\eta_1&=&\frac{2}{27}\bigg(\sqrt{4M^2-3a^2}-2M\bigg)^2
\bigg(\sqrt{4M^2-3a^2}+M\bigg), \nonumber\\ \eta_2&=&-\frac{2}{27}\bigg(\sqrt{4M^2-3a^2}+2M\bigg)^2
\bigg(\sqrt{4M^2-3a^2}-M\bigg),
\end{eqnarray}
which are functions of the mass $M$ and the rotation parameter $a$ of the spacetime. For the cases $\eta<\eta_2<0$ or $|a|>\frac{2\sqrt{3}M}{3}$ with $\eta<0$, we one can find that $r^{1}_{H}$ is negative and the roots $r^{2}_{H}$, $r^{3}_{H}$ are imaginary, which means that
there is no horizon and then the spacetime (\ref{metric1}) becomes a naked singularity. For the cases $\eta>\eta_1$ or $0<\eta<\eta_2$ or $|a|>\frac{2\sqrt{3}M}{3}$ with $\eta>0$,  the spacetime possesses a single horizon since there exist only a positive root among the three roots ($r^{1}_{H}$, $r^{2}_{H}$, $r^{3}_{H}$) in these cases. As $\eta_2\leq\eta\leq0$ or $\eta=\eta_1$,  there exist two horizons as in the usual non-extremal Kerr black hole spacetime. Especially, one can find that the one can find all of the three roots ($r^{1}_{H}$, $r^{2}_{H}$, $r^{3}_{H}$) are positive in the case $0<\eta<\eta_1$ with the negative $\eta_2$ or in the case $\eta_2<\eta<\eta_1$ with the positive $\eta_2$, which means that
black hole spacetime (\ref{metric1}) has three horizons in these cases.  These properties indicate that the presence of the deformation parameter extends the allowed range of the rotation parameter $a$ and  changes the spacetime structure in the strong field region \cite{W6}.
\begin{figure}\label{fig1}
\begin{center}
\includegraphics[width=6cm]{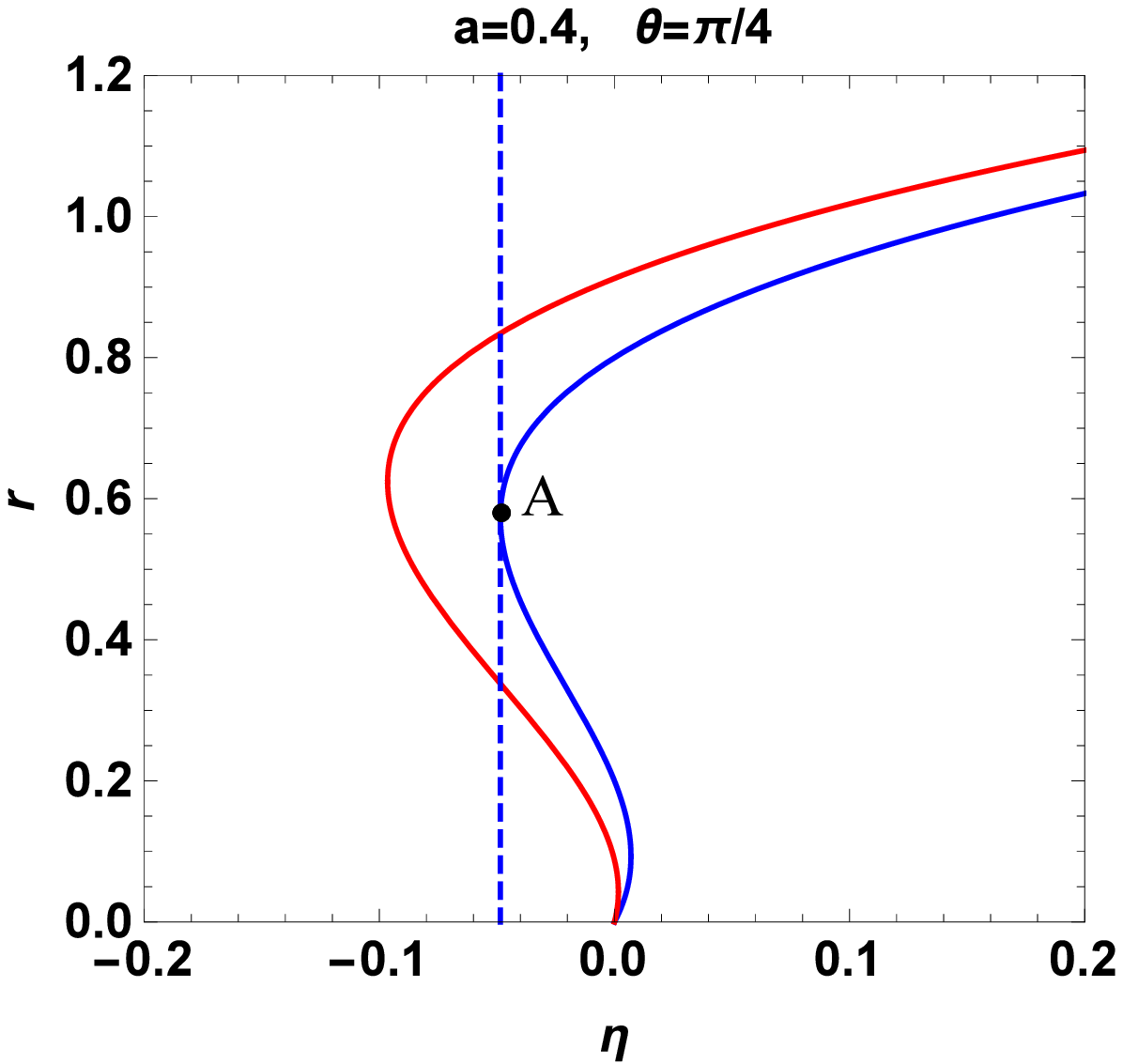}\;\;\;\;
\includegraphics[width=6cm]{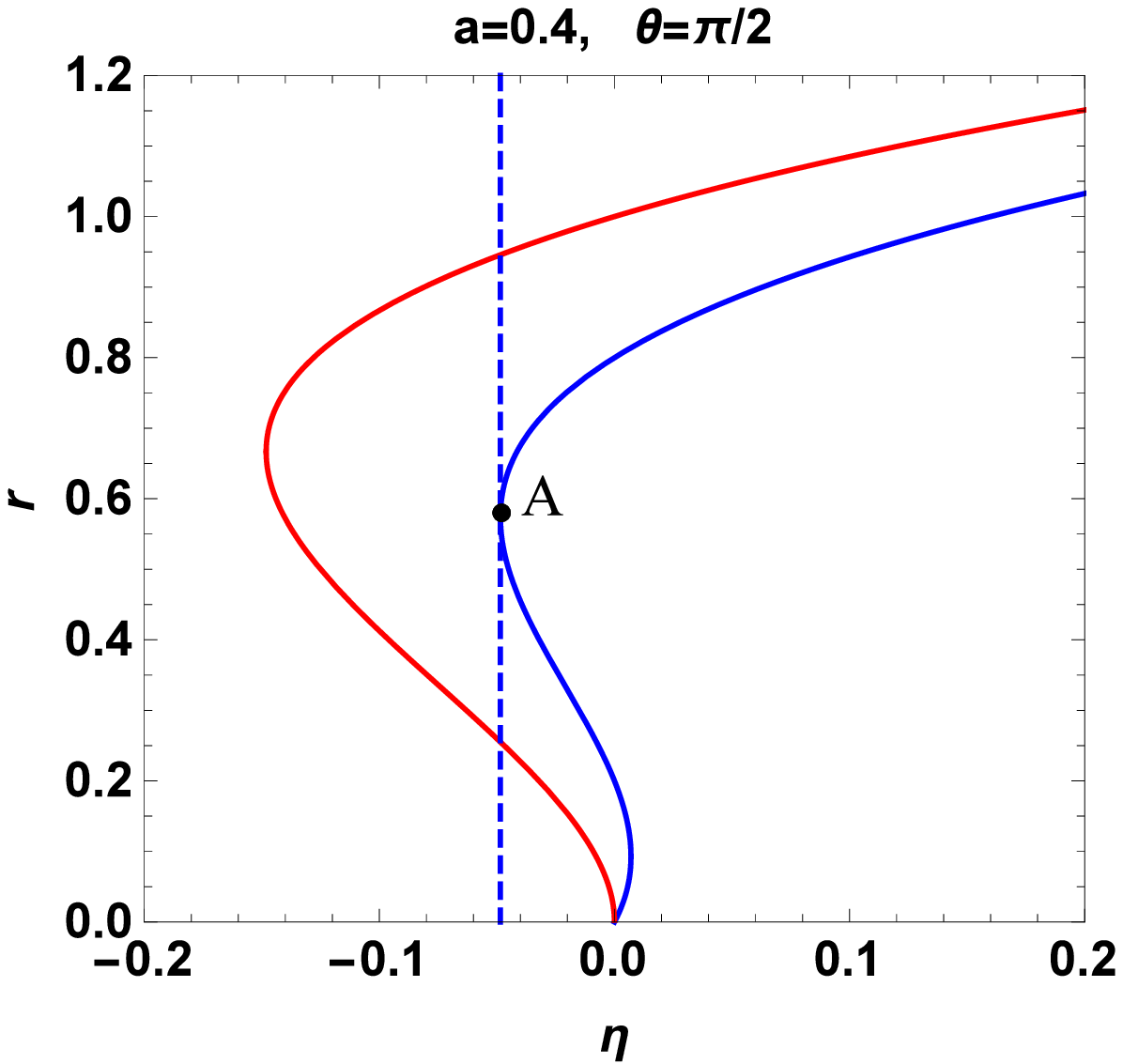}\\
\includegraphics[width=6cm]{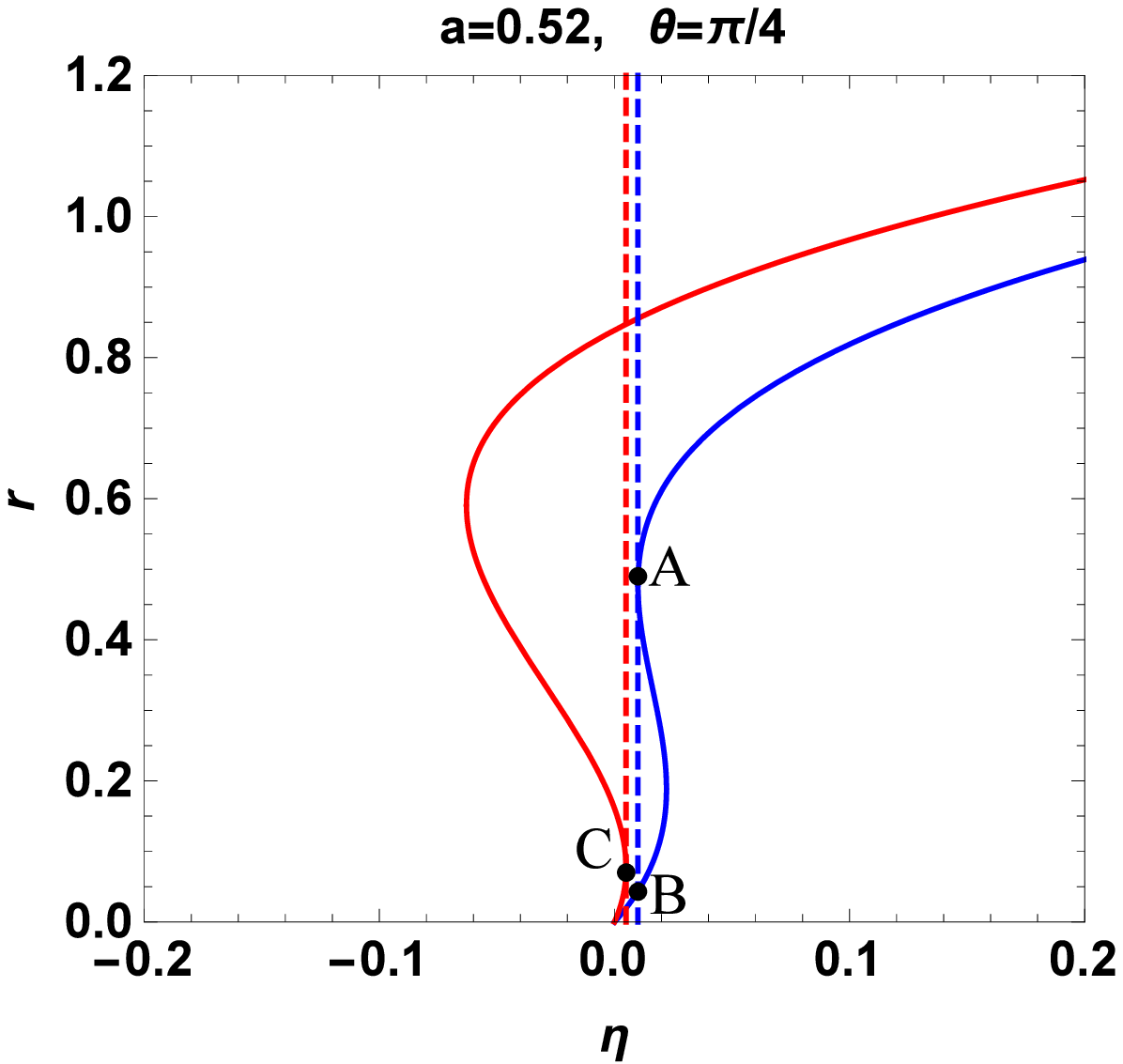}\;\;\;\;
\includegraphics[width=6cm]{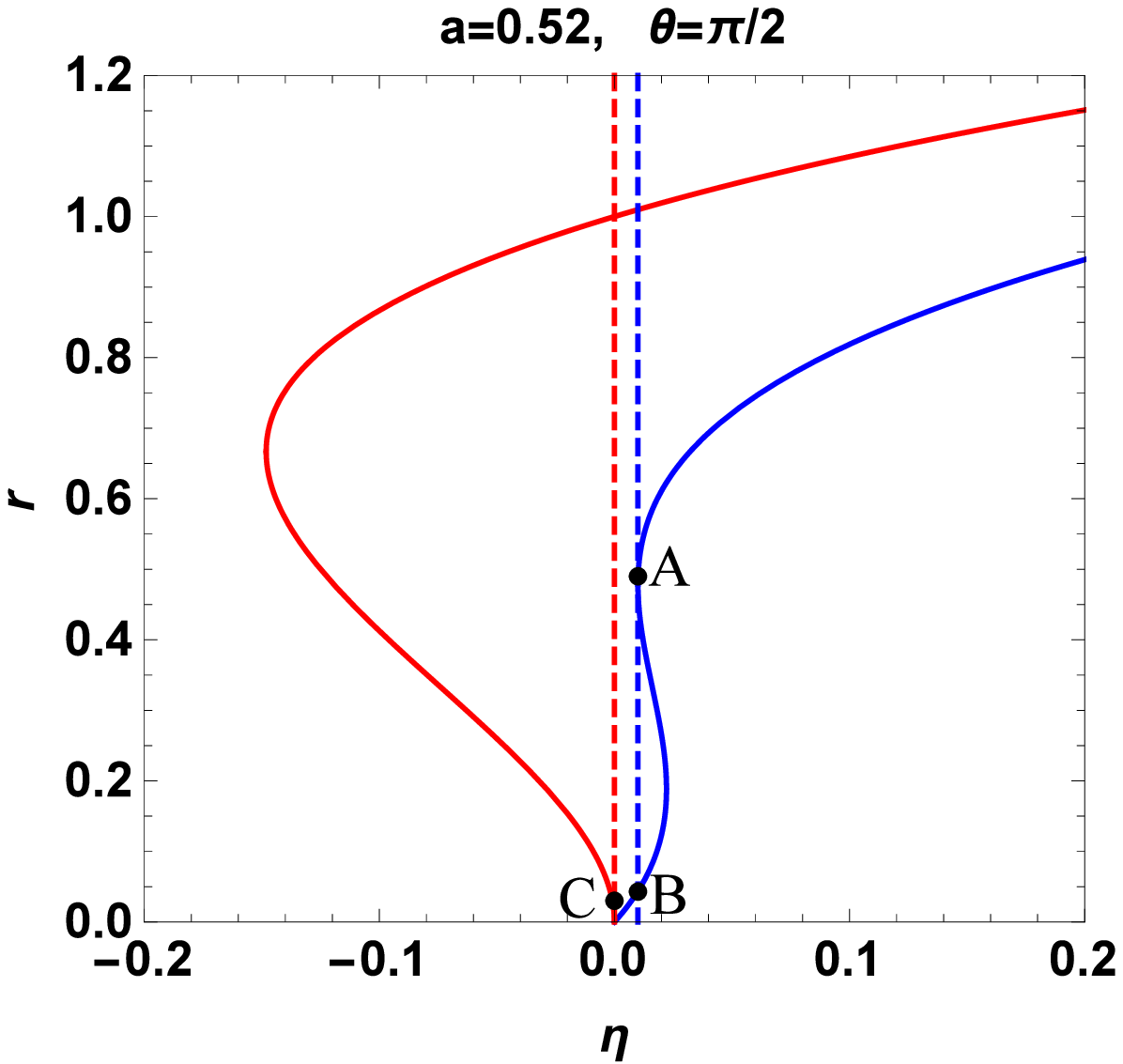}\\
\includegraphics[width=6cm]{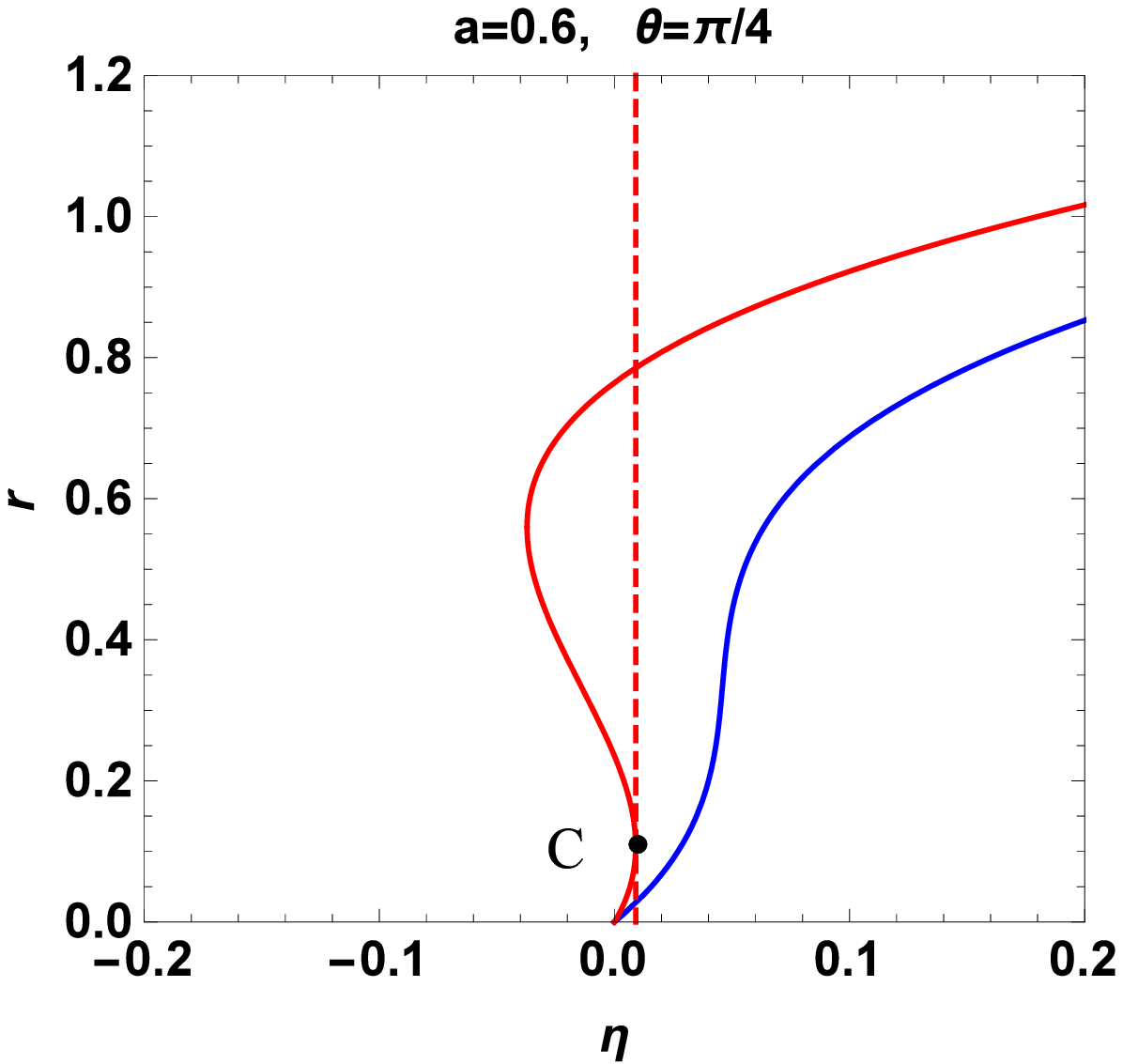}\;\;\;\;
\includegraphics[width=6cm]{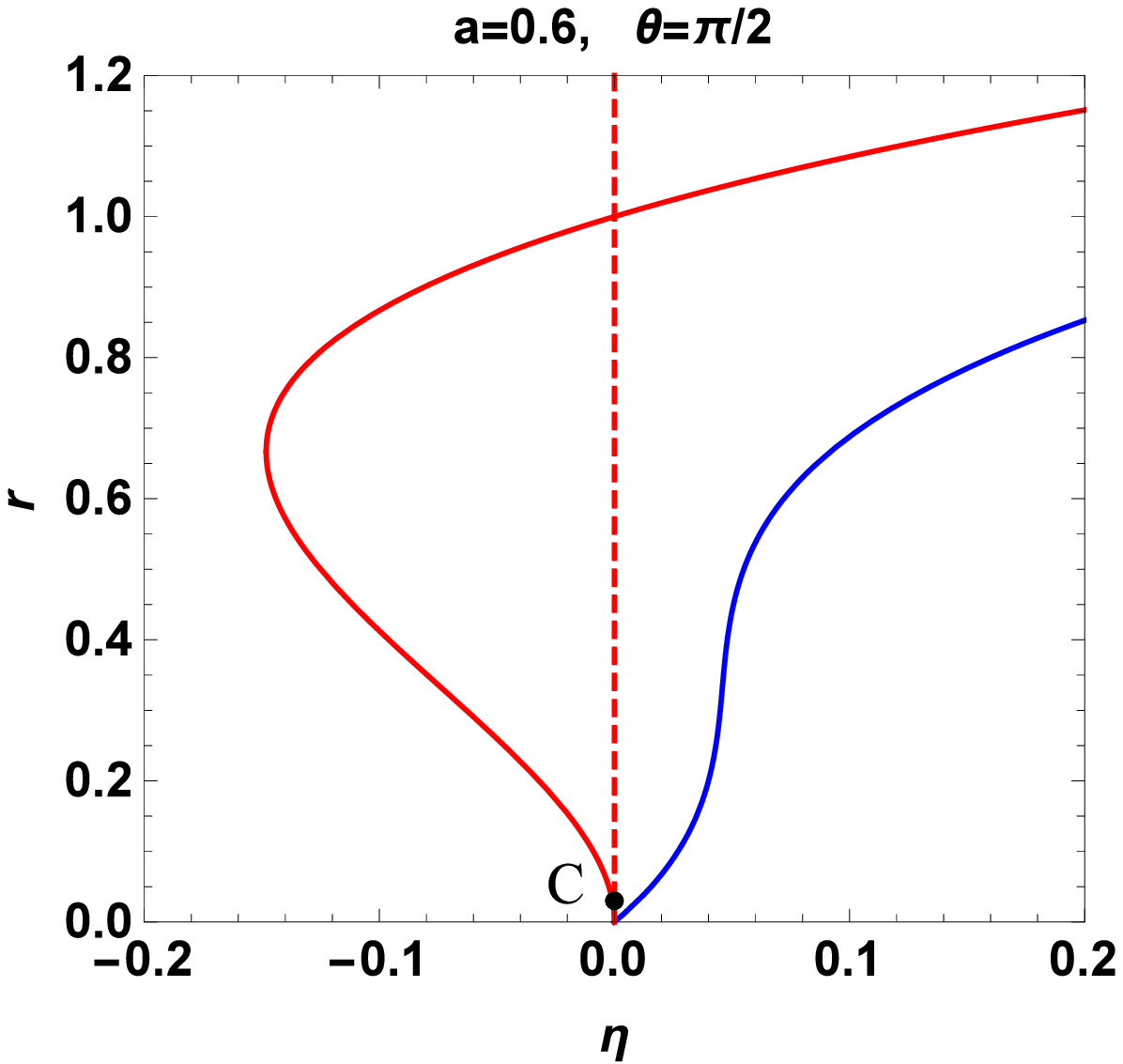}
\caption{The variation of the event horizon radius and the infinite redshift surface with the deformation parameter $\eta$ in the Konoplya-Zhidenko rotating  non-Kerr spacetime. The blue, red curves correspond to event horizon radius and the infinite redshift surface, respectively. The blue dashed line and the red dashed line correspond to the curves $\eta=\eta_2$ and $\eta=\eta_3$, respectively.  Here, we take $2M=1$. }
\end{center}
\end{figure}

For a Konoplya-Zhidenko rotating non-Kerr black hole, the infinite redshift surface is defined by the equation
\begin{eqnarray}
r^3-2Mr^2+a^2r\cos^2\theta-\eta=0.
\end{eqnarray}
Similarly, one can obtain three roots of the above equation, i.e.,
\begin{eqnarray}
r^{1}_{\infty}&=&\frac{1}{3}\bigg[2M+\frac{2^{1/3}(4M^2- 3a^2\cos^2\theta)}{\mathcal{B}^{1/3}}+
\frac{\mathcal{B}^{1/3}}{2^{1/3}}\bigg],\nonumber\\
r^{2}_{\infty}&=&\frac{1}{3}\bigg[2M-\frac{(1+\sqrt{3}i)(4M^2- 3a^2\cos^2\theta)}{2^{2/3}\mathcal{B}^{1/3}}-
\frac{(1-\sqrt{3}i)\mathcal{B}^{1/3}}{2^{4/3}}\bigg],\nonumber\\
r^{3}_{\infty}&=&\frac{1}{3}\bigg[2M-\frac{(1-\sqrt{3}i)(4M^2- 3a^2\cos^2\theta)}{2^{2/3}\mathcal{B}^{1/3}}-
\frac{(1+\sqrt{3}i)\mathcal{B}^{1/3}}{2^{4/3}}\bigg],
\end{eqnarray}
with
\begin{eqnarray}
\mathcal{B}=16M^3+27\eta-18 a^2M\cos^2\theta + \sqrt{(16M^3+27\eta-18 a^2M\cos^2\theta)^2-4(4M^2-3a^2 \cos^2\theta)^3}.
\end{eqnarray}
\begin{figure}
\begin{center}
\includegraphics[width=3.82cm]{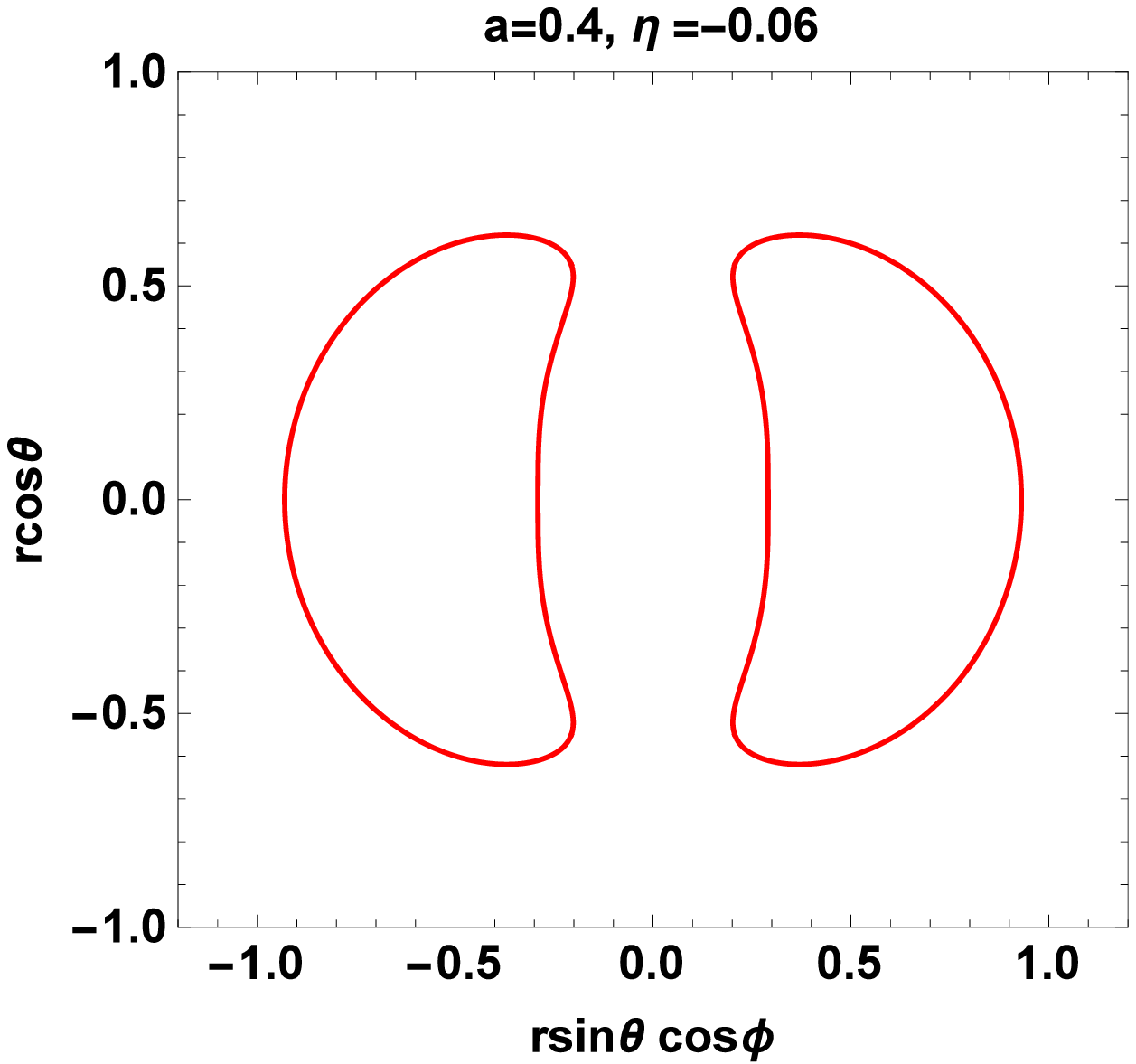}\;\;\;\;
\includegraphics[width=3.82cm]{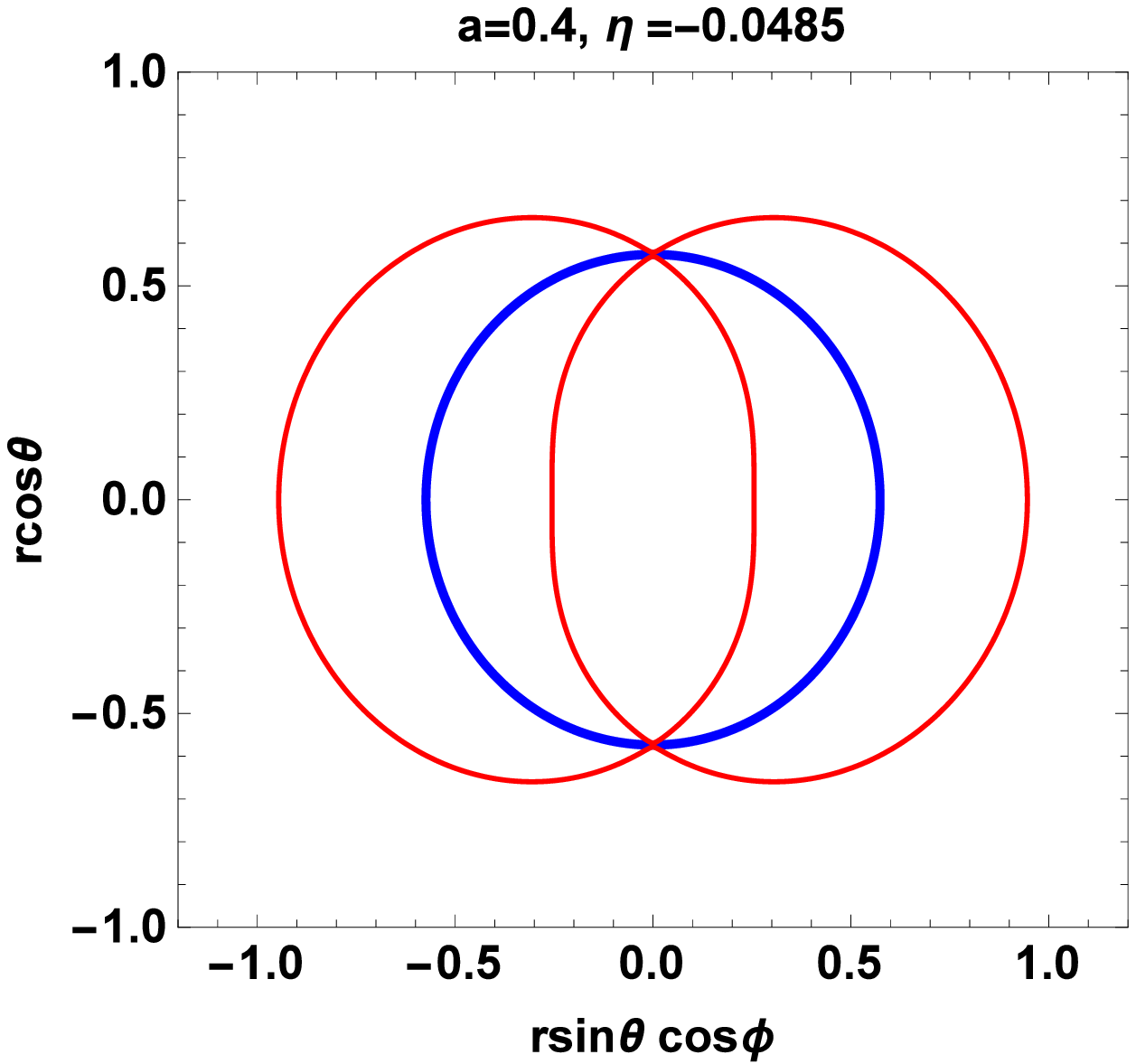}\;\;\;\;
\includegraphics[width=3.82cm]{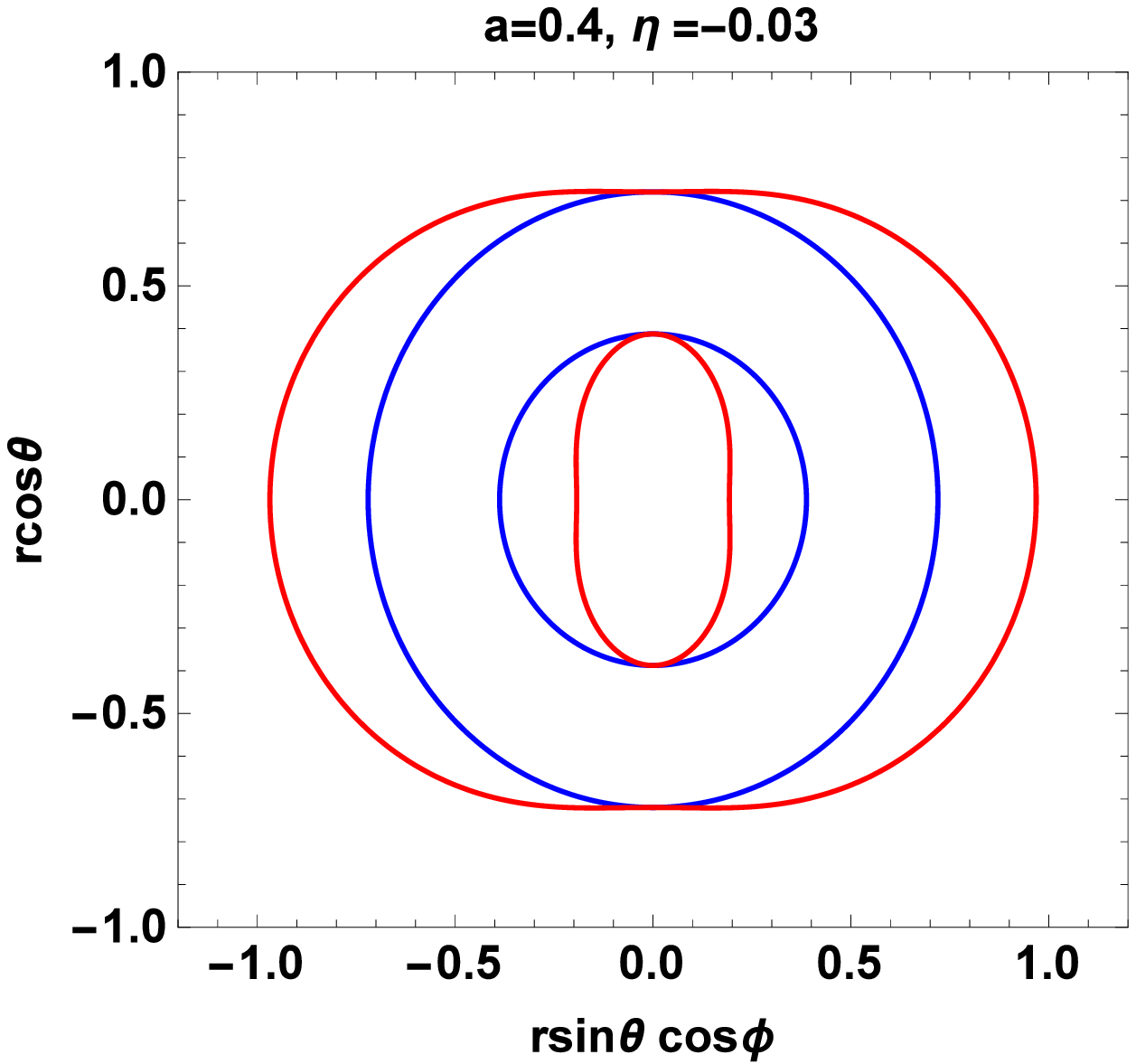}\;\;\;\;
\includegraphics[width=3.82cm]{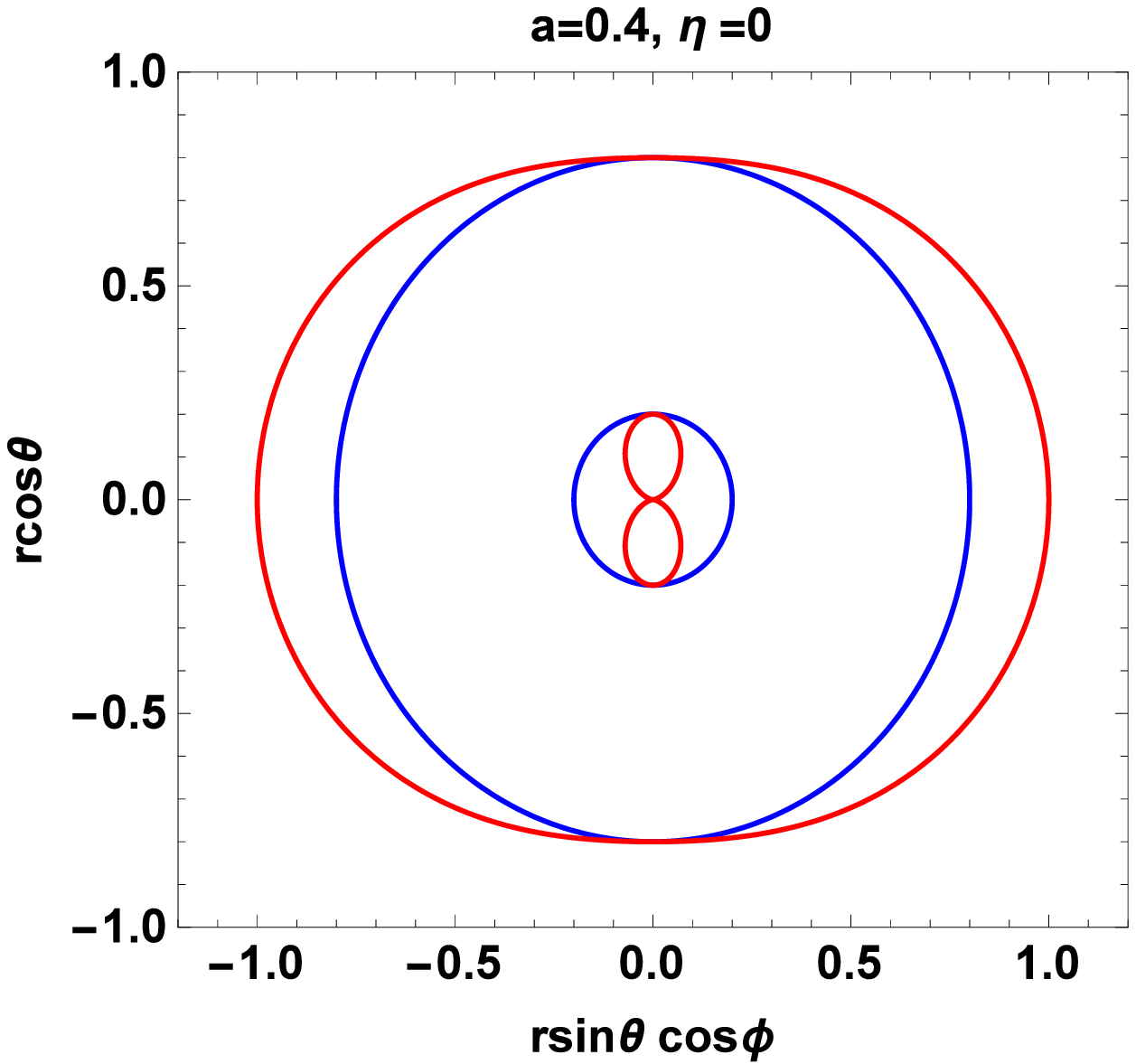}\\
\includegraphics[width=3.82cm]{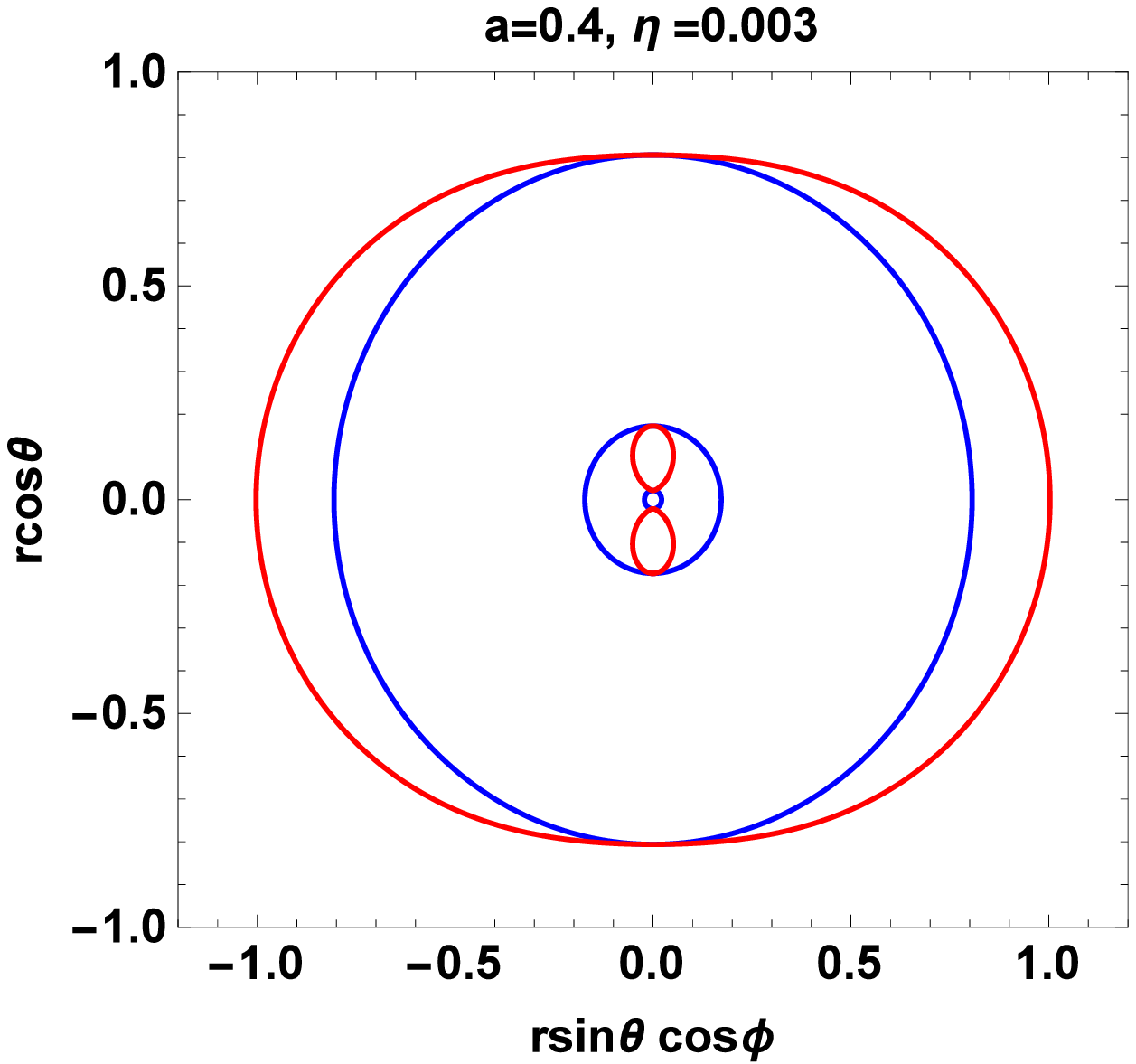}\;\;\;\;
\includegraphics[width=3.82cm]{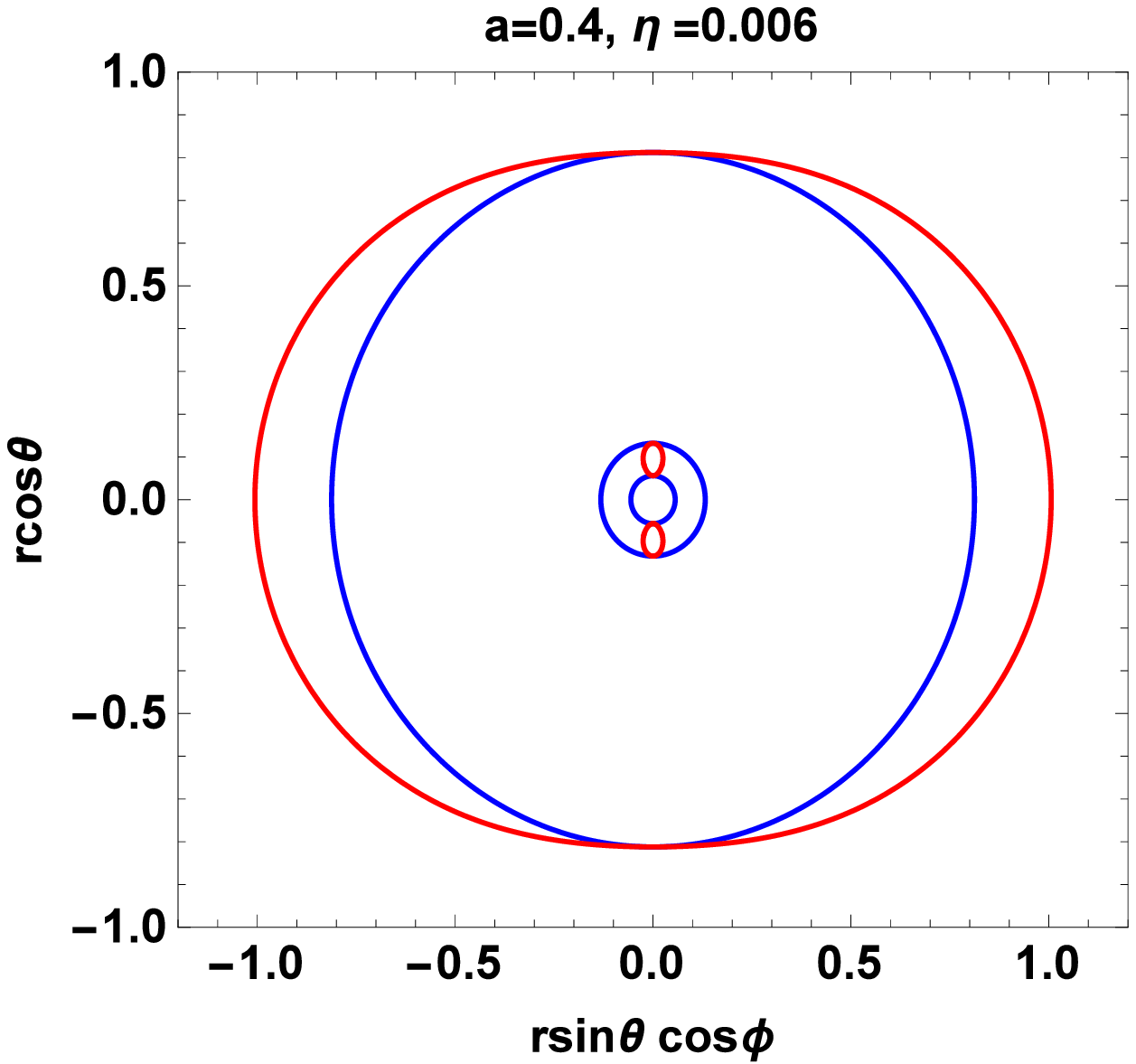}\;\;\;\;
\includegraphics[width=3.82cm]{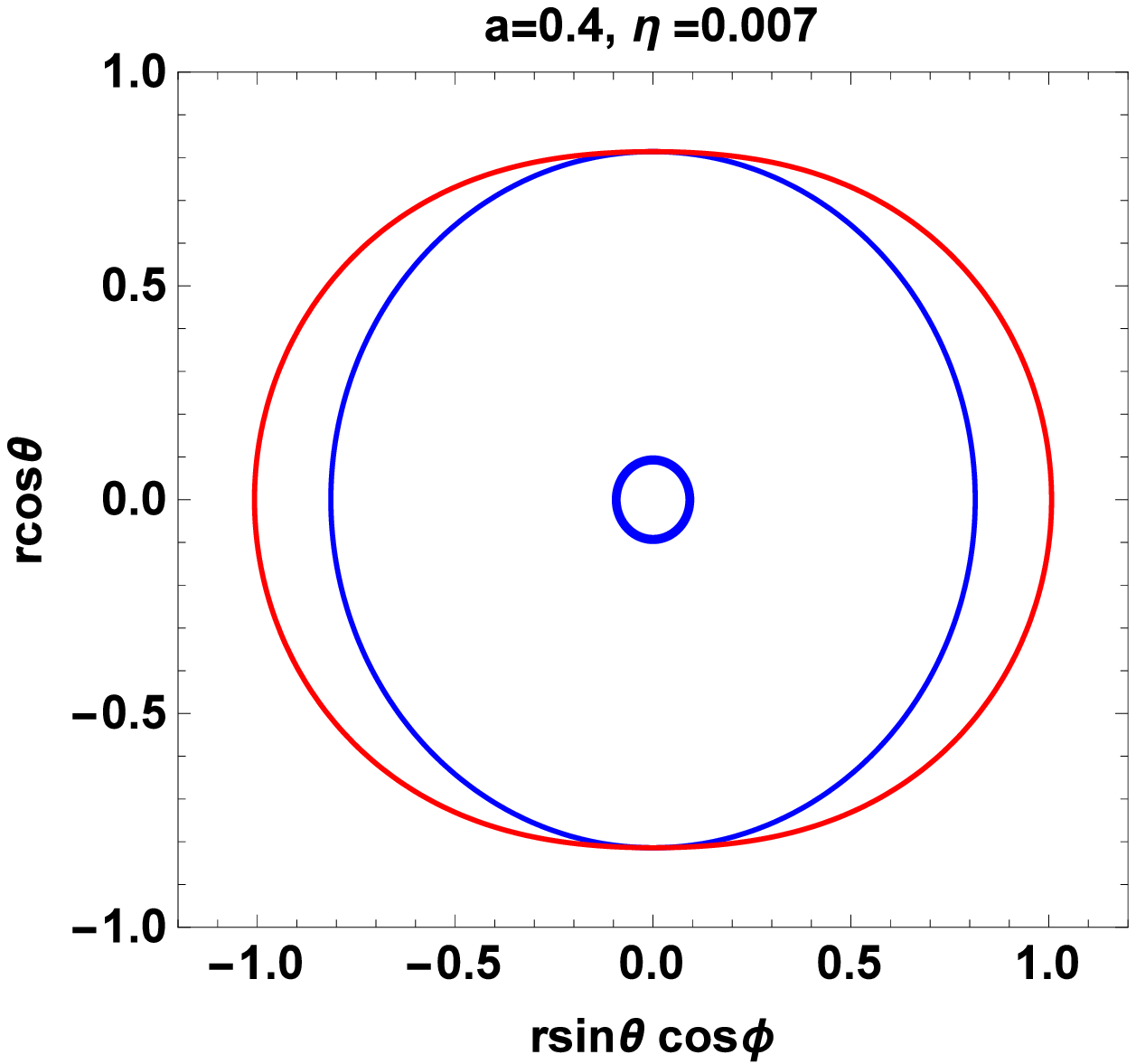}\;\;\;\;
\includegraphics[width=3.82cm]{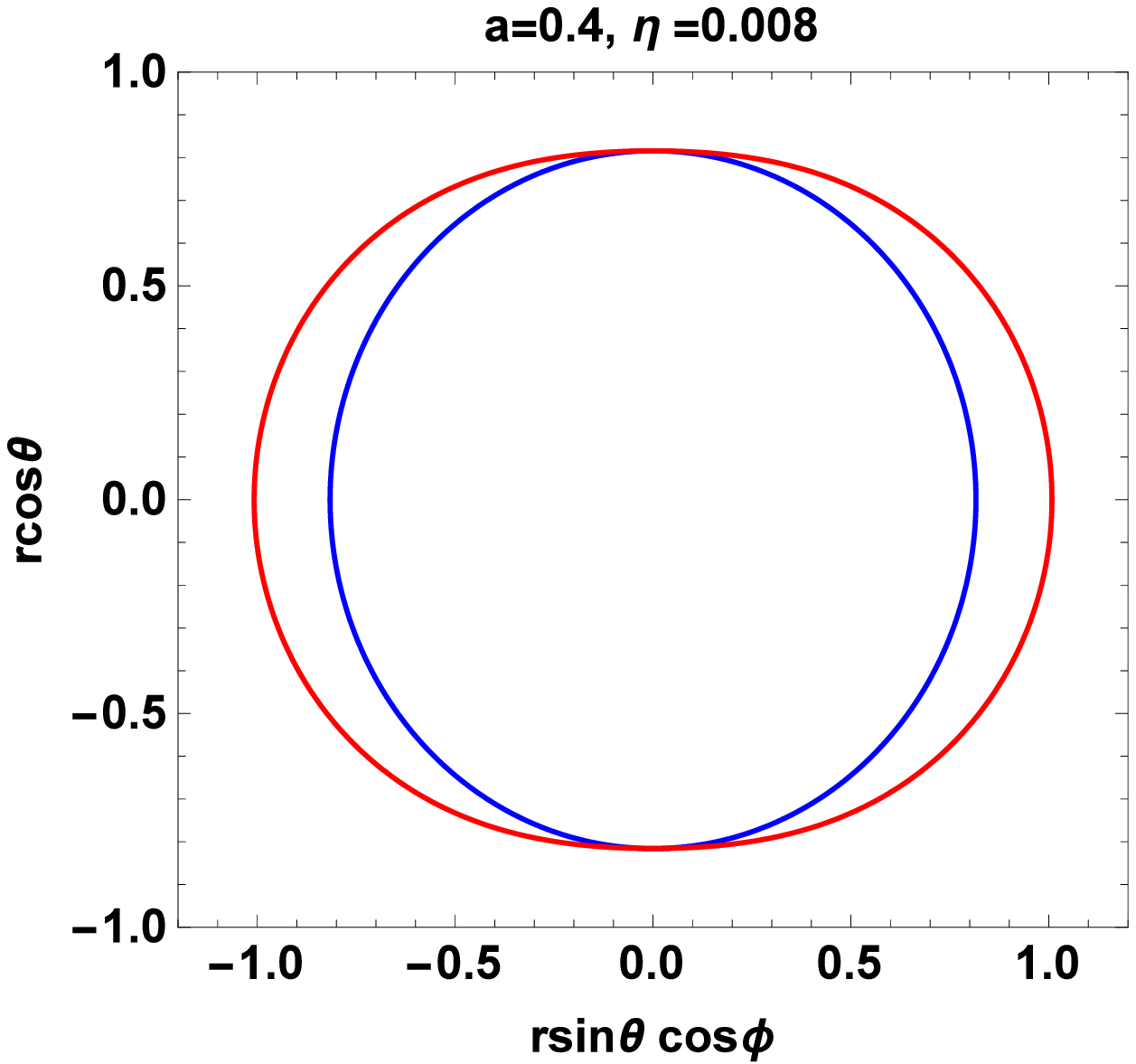}
\caption{The Variation of the shape on the \textit{xz}-plane of the ergosphere with the deformation parameter $\eta$ in the Konoplya-Zhidenko rotating  non-Kerr spacetime with fixed $a=0.4$. The red and the blue lines correspond to the infinite redshift surfaces and the horizons, respectively. Here, we take $M=0.5$. }
\end{center}
\end{figure}
\begin{figure}
\begin{center}
\includegraphics[width=3.82cm]{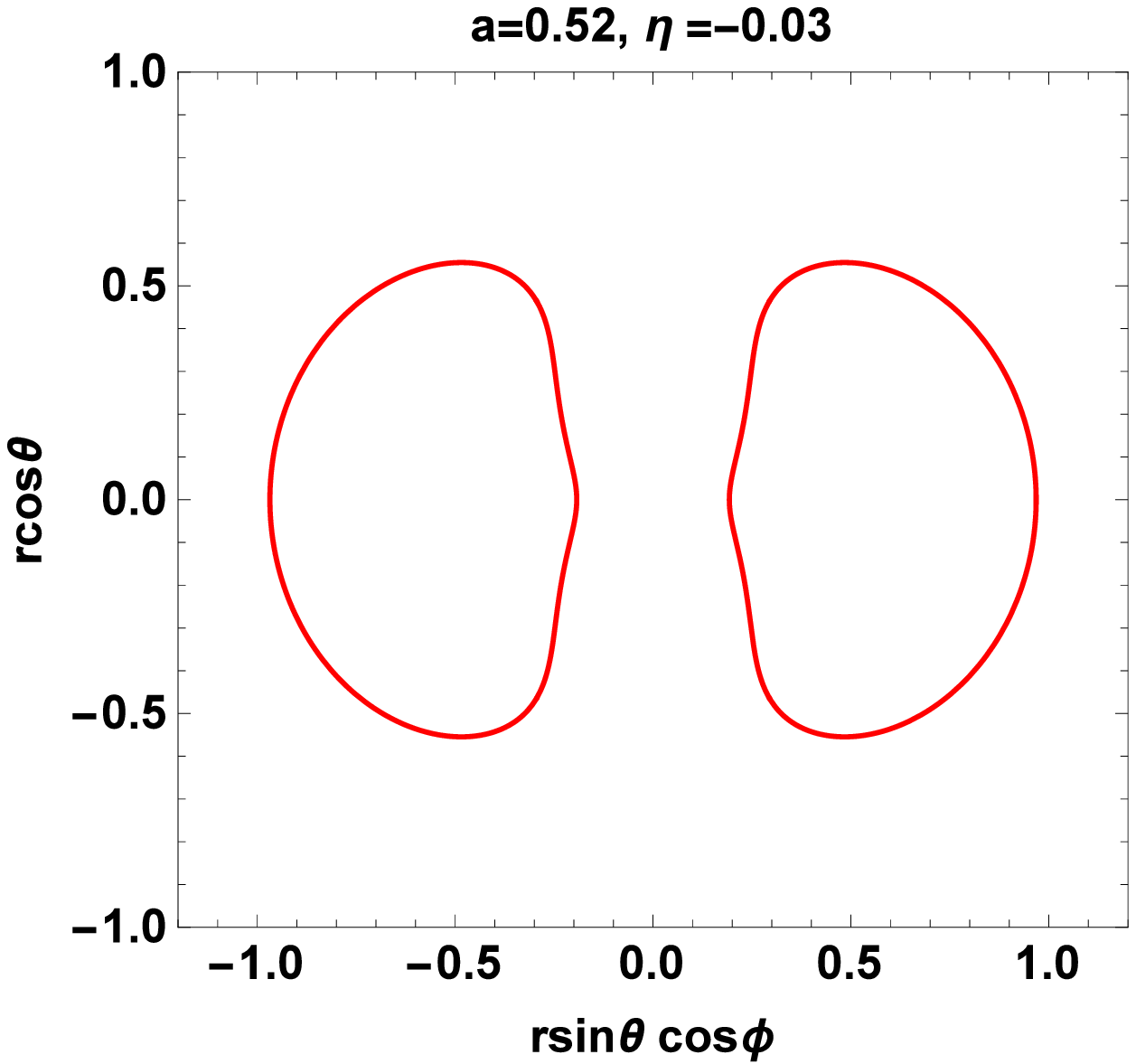}\;\;\;\;
\includegraphics[width=3.82cm]{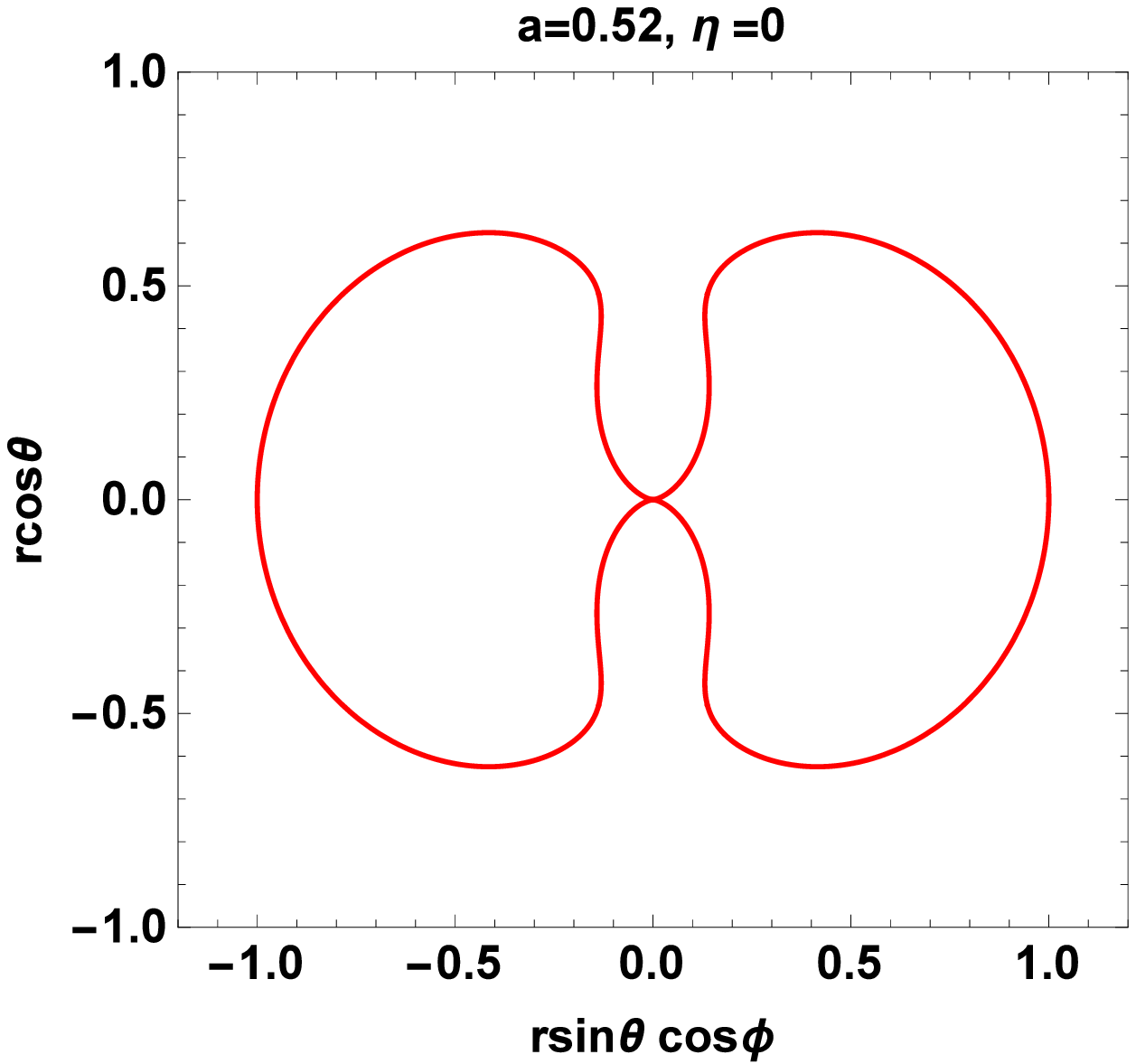}\;\;\;\;
\includegraphics[width=3.82cm]{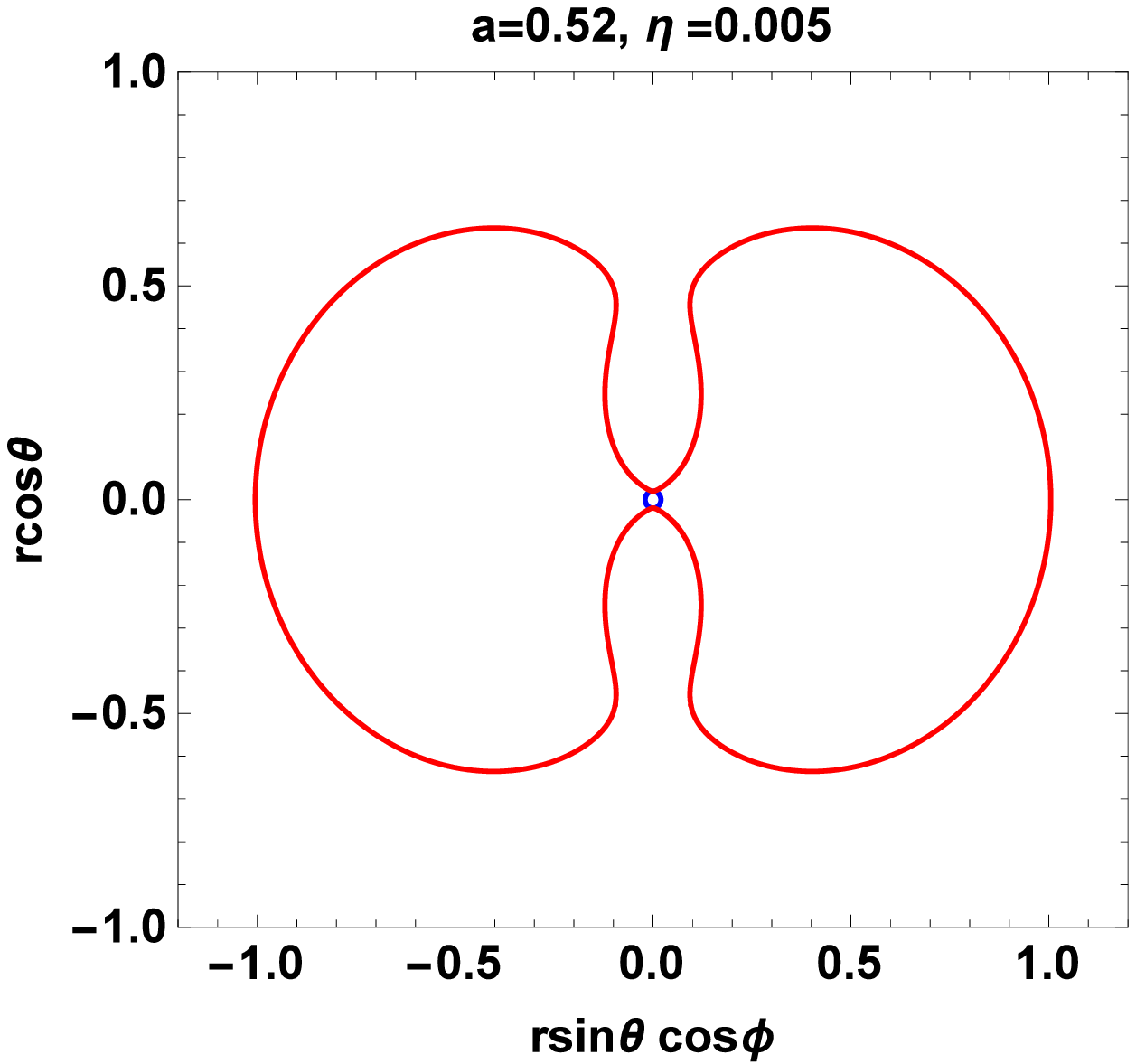}\;\;\;\;
\includegraphics[width=3.82cm]{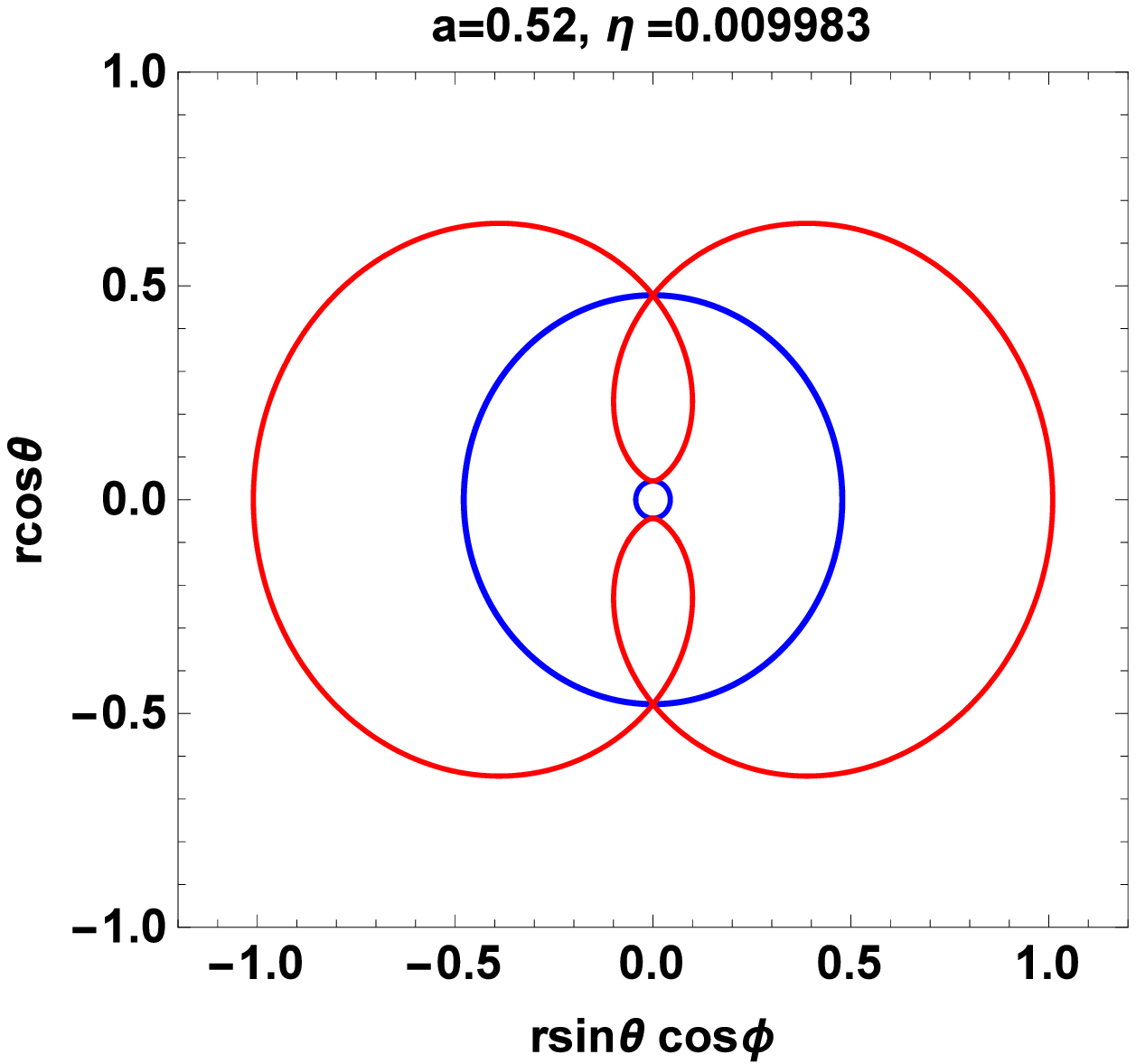}\\
\includegraphics[width=3.82cm]{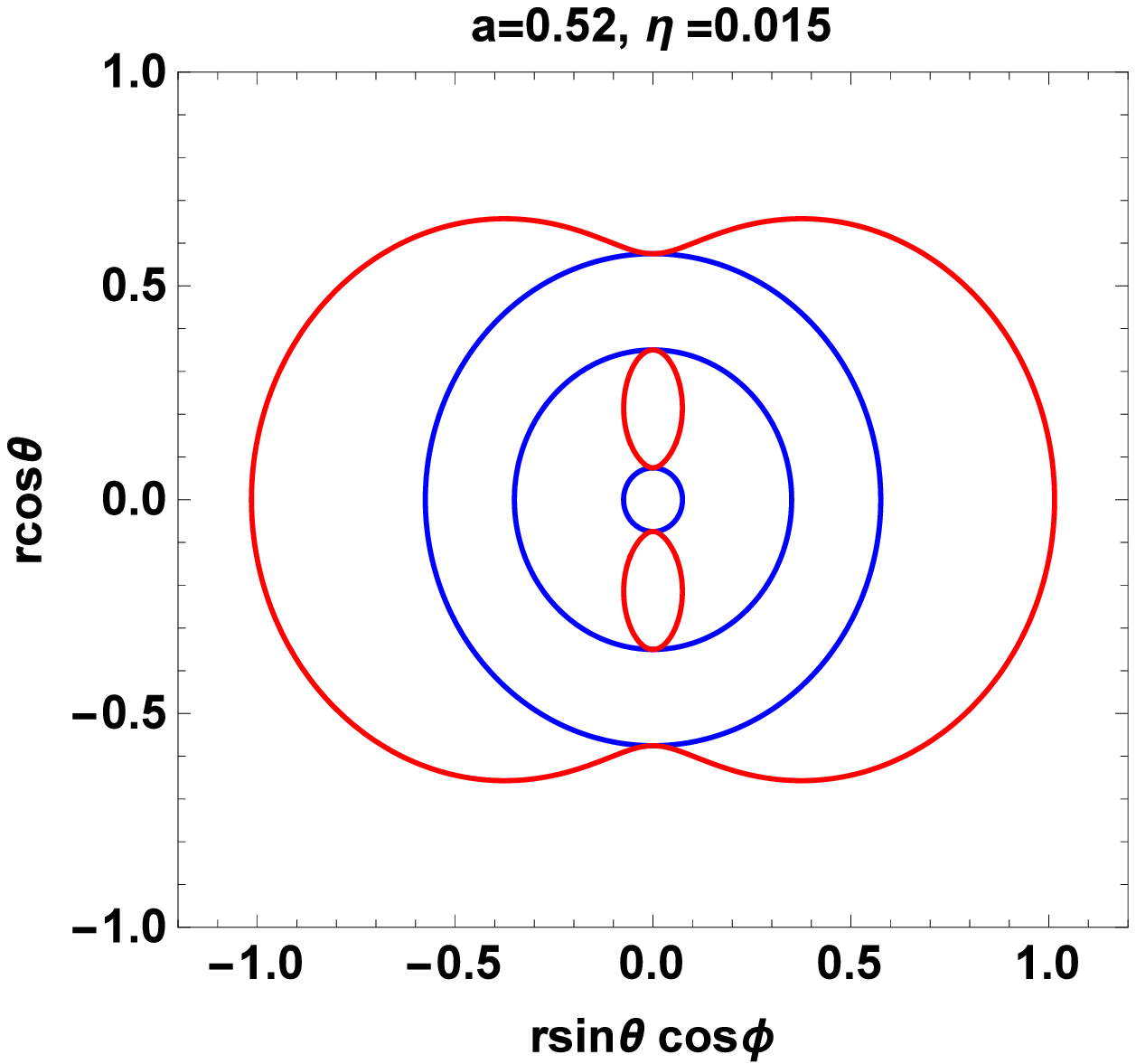}\;\;\;\;
\includegraphics[width=3.82cm]{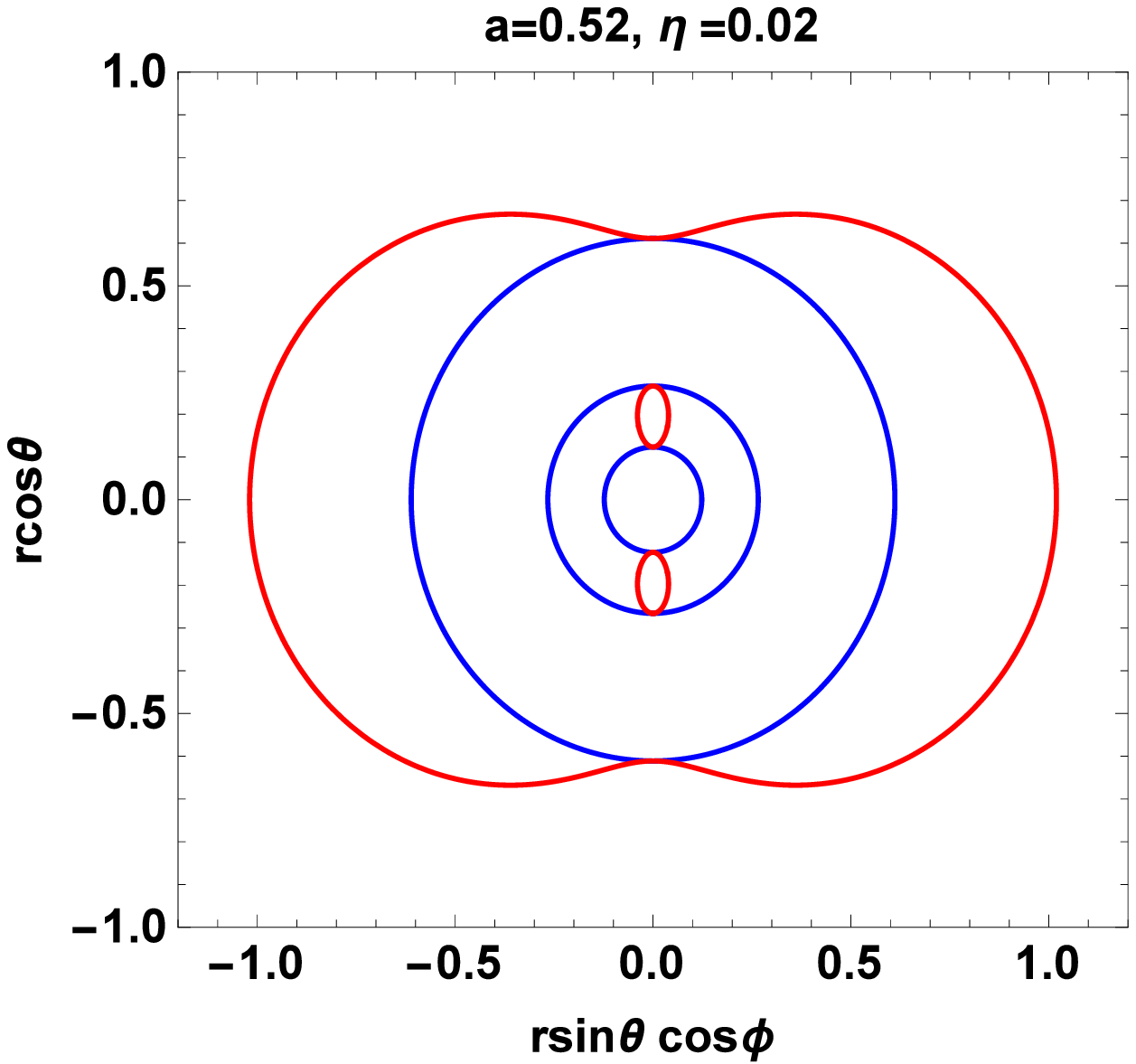}\;\;\;\;
\includegraphics[width=3.82cm]{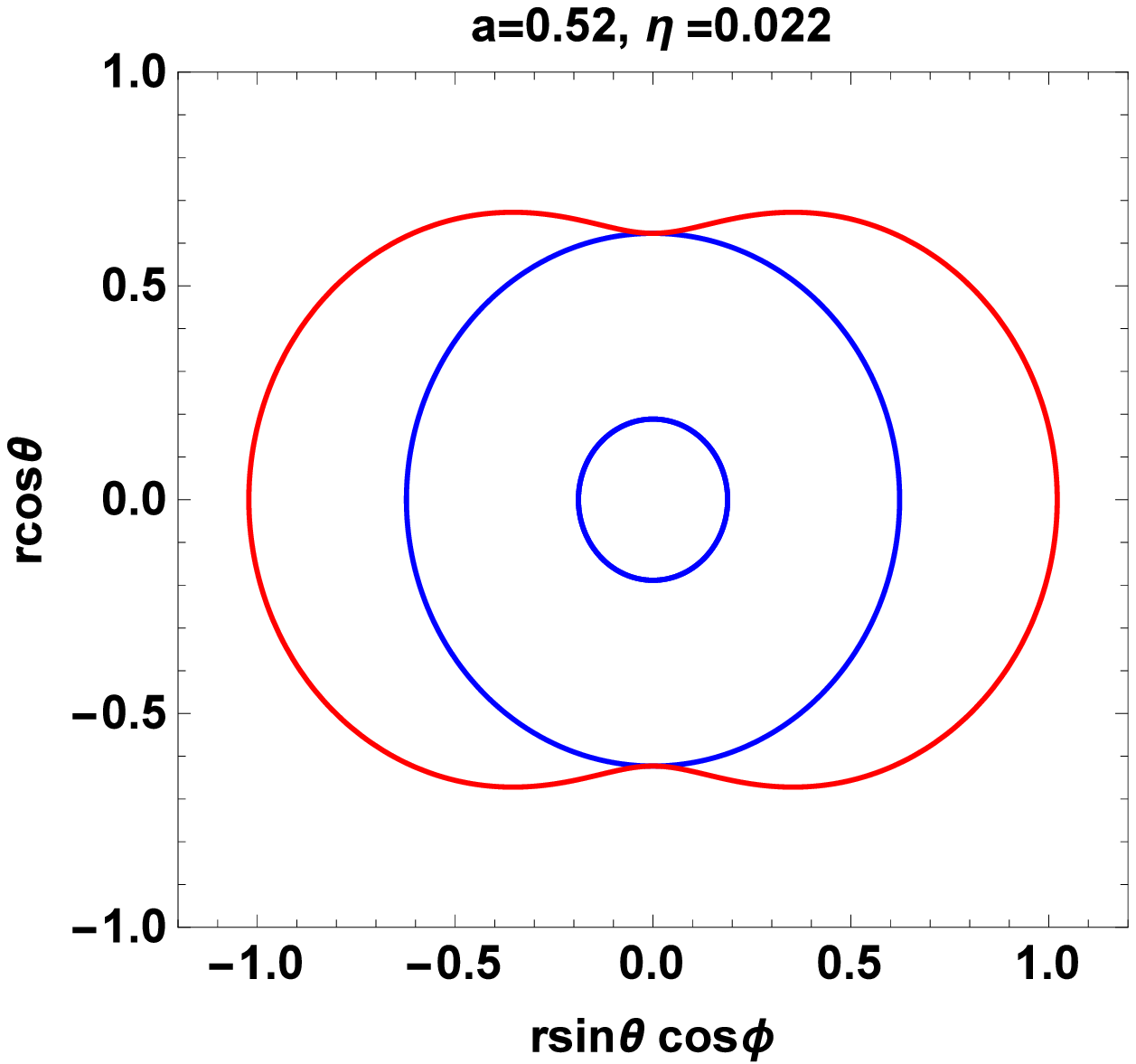}\;\;\;\;
\includegraphics[width=3.82cm]{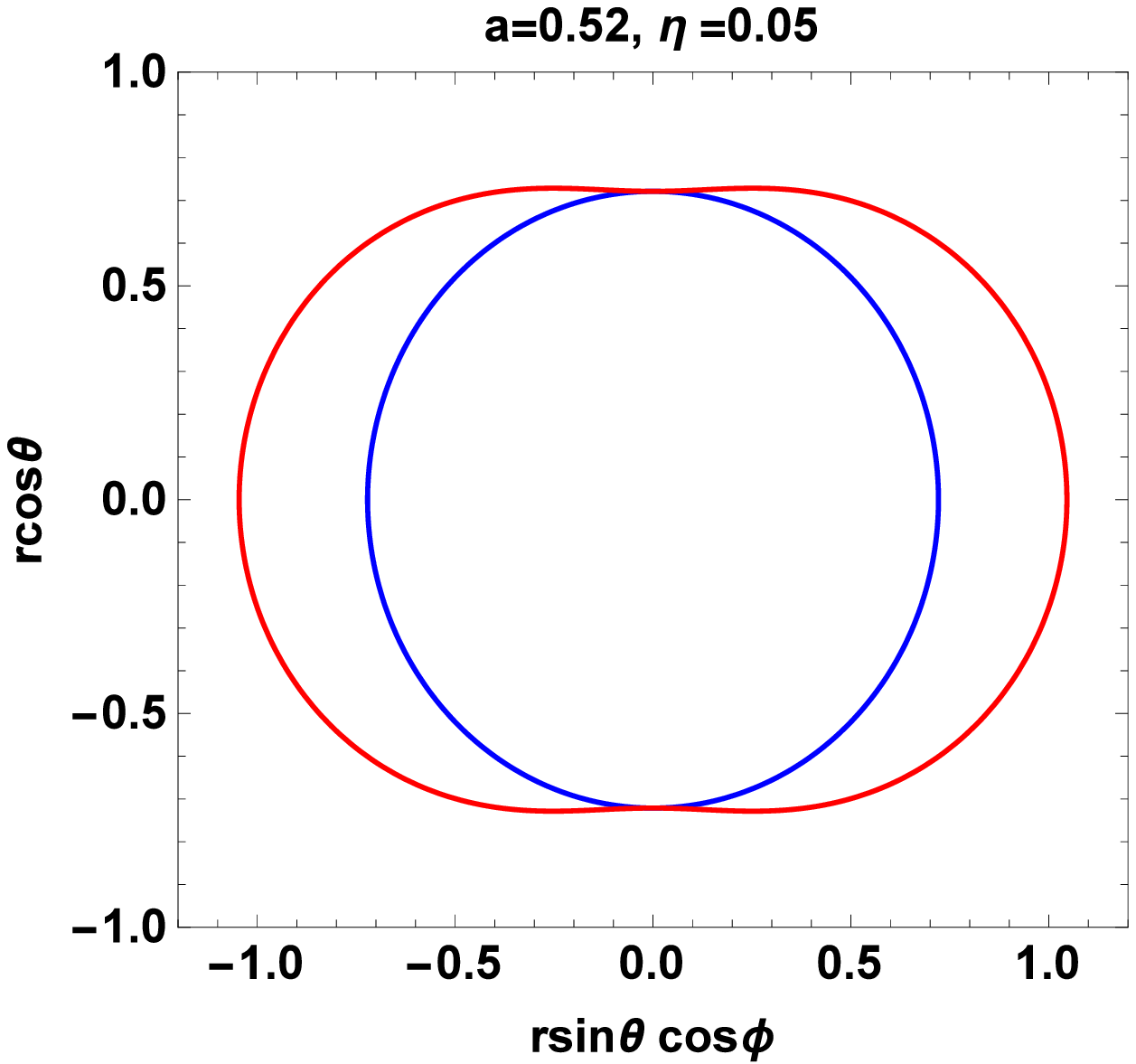}
\caption{The Variation of the shape on the \textit{xz}-plane of the ergosphere with the deformation parameter $\eta$ in the Konoplya-Zhidenko rotating  non-Kerr spacetime with fixed $a=0.52$. The red and the blue lines correspond to the infinite redshift surfaces and the horizons, respectively. Here, we take $M=0.5$. }
\end{center}
\end{figure}
\begin{figure}
\begin{center}
\includegraphics[width=3.82cm]{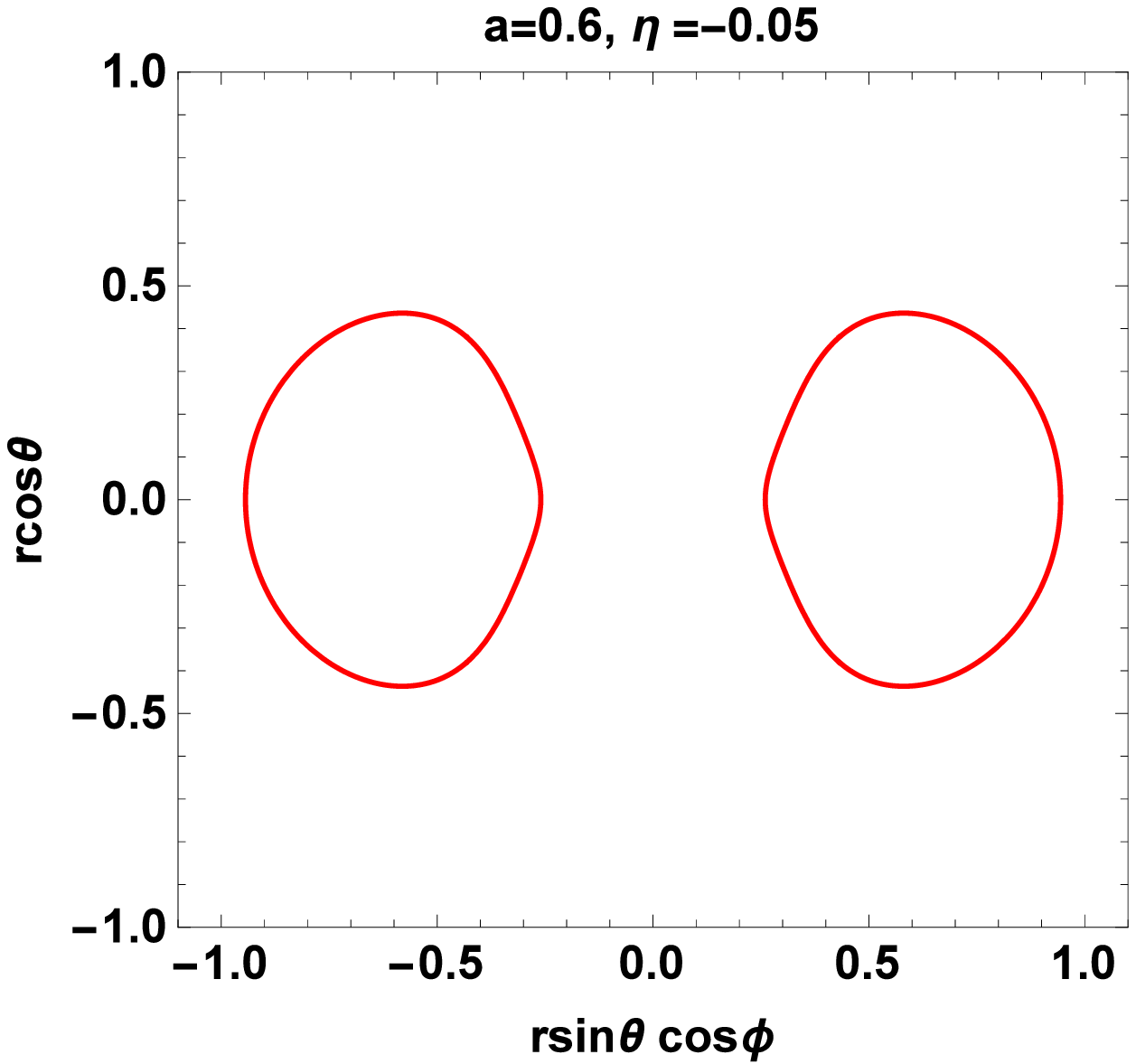}\;\;\;\;
\includegraphics[width=3.82cm]{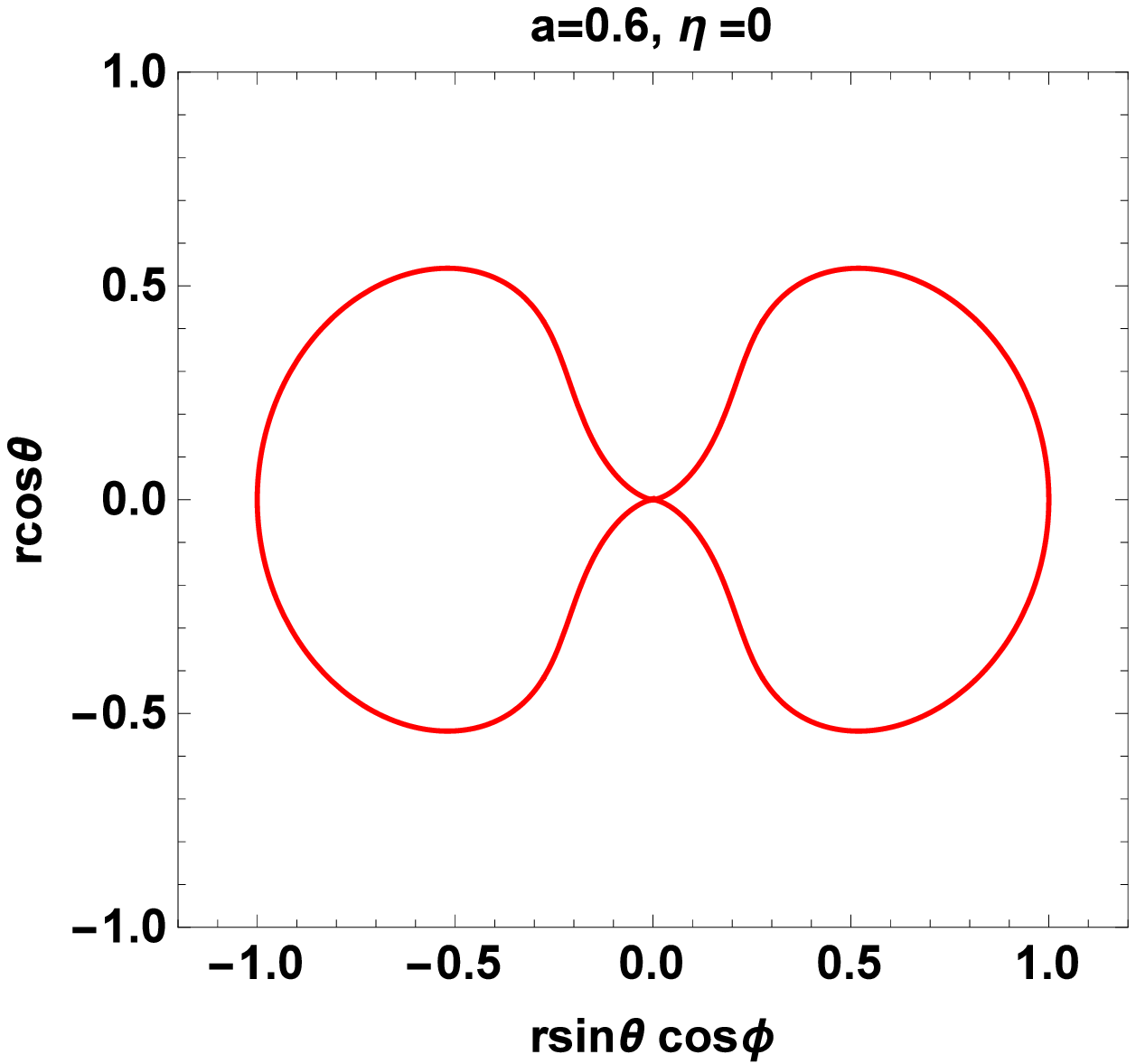}\;\;\;\;
\includegraphics[width=3.82cm]{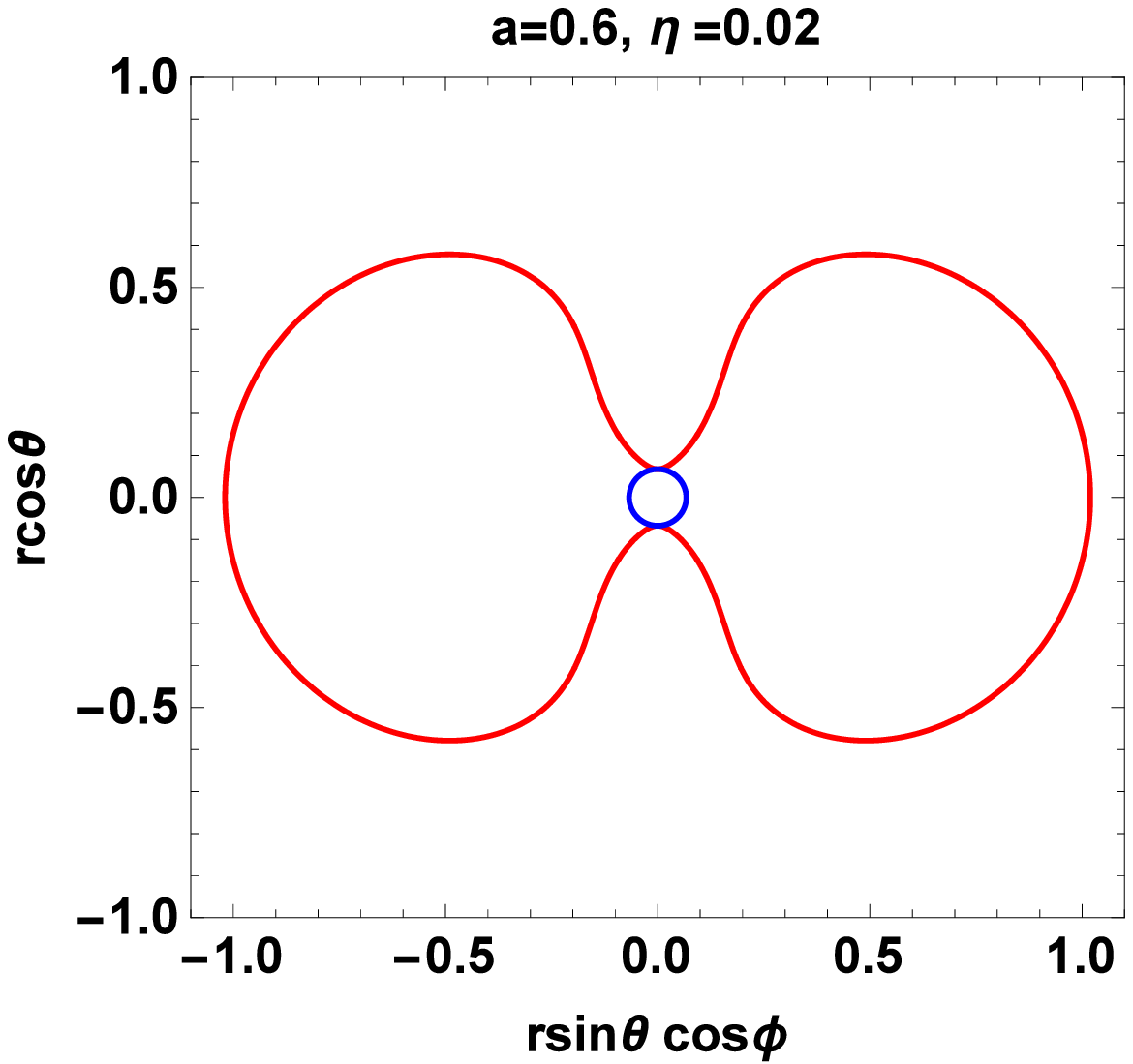}\;\;\;\;
\includegraphics[width=3.82cm]{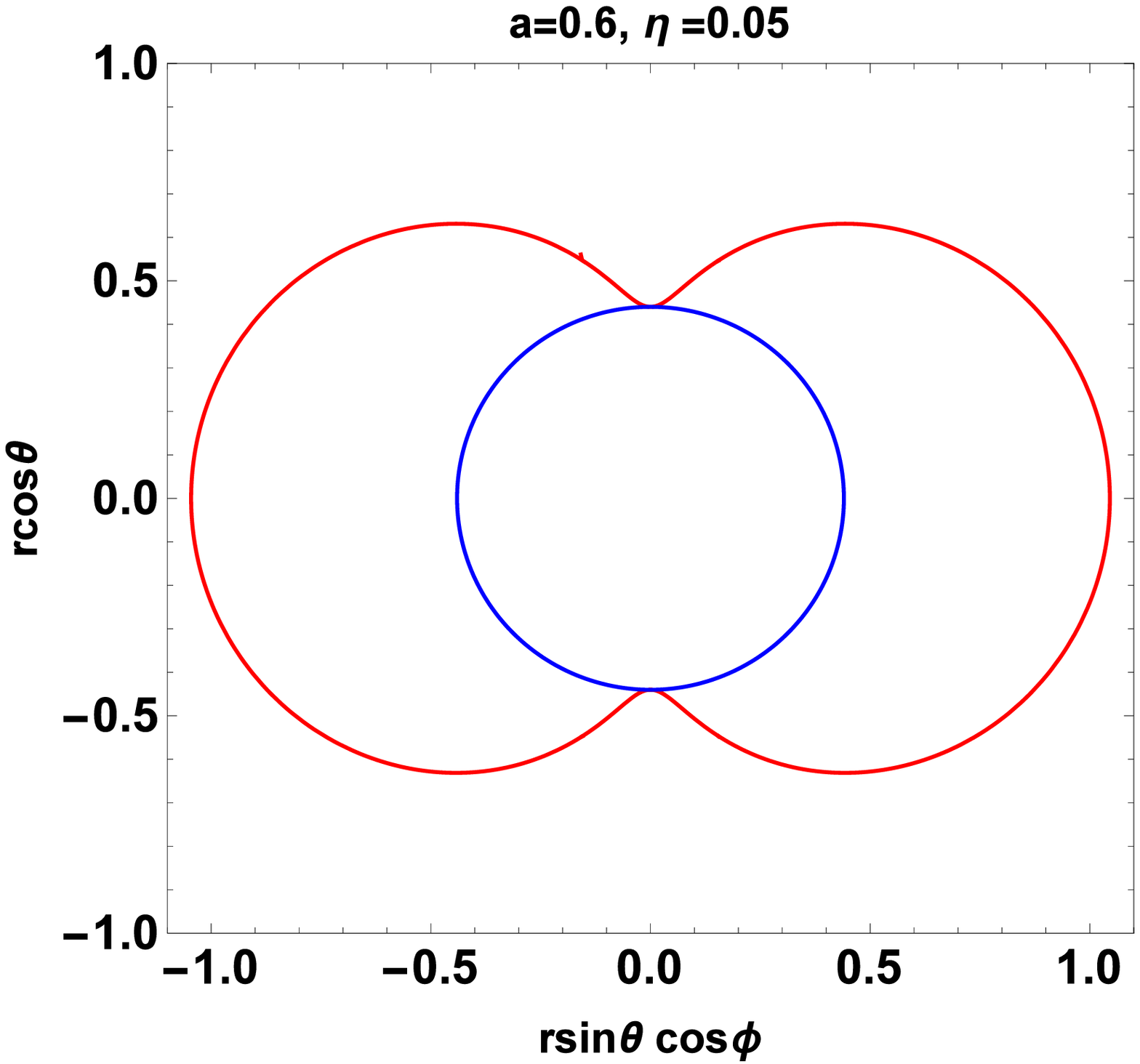}
\caption{The Variation of the shape on the \textit{xz}-plane of the ergosphere with the deformation parameter $\eta$ in the Konoplya-Zhidenko rotating  non-Kerr spacetime with fixed $a=0.6$. The red and the blue lines correspond to the infinite redshift surfaces and the horizons, respectively. Here, we take $M=0.5$. }
\end{center}
\end{figure}
It is easy to find that the outer infinite redshift surface is located at $r=r^{1}_{\infty}$. The ergosphere is the region bounded by the outer event horizon and the outer infinite redshift surface, which is very important for the rotation energy extraction from a Kerr-like black hole \cite{W1}. Comparing with the case of a Kerr black hole, we find that the region of ergosphere becomes more complicated in the Konoplya-Zhidenko rotating non-Kerr spacetime due to the deformation parameter $\eta$. In Fig.(1), we present the variation of the event horizon radius and the infinite redshift surface with the deformation parameter $\eta$. There exists a turning point $A$ for the event horizon surface at which $\eta=\eta_2$.  For the case with $a<M$, the ergosphere is determined by the difference between $r^{1}_{\infty}$ and $r^{1}_{H}$ as $\eta\geq\eta_2$, but as $\eta<\eta_2$ the spacetime (\ref{metric1}) is a naked singularity in which there exists no any horizon. For the case with $M<a<\frac{2\sqrt{3}M}{3}$, we find that the ergosphere is still determined by the difference between $r^{1}_{\infty}$ and $r^{1}_{H}$ as $\eta\geq\eta_2$, but it is given by the difference between $r^{1}_{\infty}$ and $r^{2}_{H}$ as $\eta_3\leq\eta<\eta_2$. The quantity $\eta_3$ is the value of $\eta$ at a turning point $C$ for the infinite redshift surface, which is given by
\begin{eqnarray}\label{etam0} \eta_3=-\frac{2}{27}\bigg(\sqrt{4M^2-3a^2\cos^2\theta}+2M\bigg)^2
\bigg(\sqrt{4M^2-3a^2\cos^2\theta}-M\bigg).
\end{eqnarray}
In this case, the width of the ergosphere undergoes a sudden change at the point $\eta=\eta_2$ since the position of the outer horizon jumps from $r=r^{1}_{H}$ at the point $A$ into $r=r^{2}_{H}$ at the point $B$, which is shown in the panels in the middle row of Fig.(1). This phenomenon does not emerge in the Johannsen-Psaltis non-Kerr black hole spacetime \cite{chen2,chen201}.
For the high-spinning case with $a>\frac{2\sqrt{3}M}{3}$, the turning point $A$ disappears since the $\eta_2$ is imaginary, and the ergosphere is determined by the difference between $r^{1}_{\infty}$ and $r^{2}_{H}$.

In Figs.(2)-(4), we present the dependence of the ergosphere on the deformation parameter $\eta$ for the different rotation parameter $a$. It is shown clearly that the ergosphere in the equatorial plane becomes thin with increase of the deformation parameter $\eta$, which is different from that in the Johannsen-Psaltis non-Kerr black hole spacetime in which the ergosphere become thick with the deformation parameter \cite{chen2,chen201}. It could be understandable since  there exists the deformation difference between two non-Kerr black hole metrics. Moreover, we also note that  as $\eta<\eta_2$ the horizon disappears  and the shape of the infinite redshift surface becomes toroidal. Especially, in the case with $a>M$, we can find that as the positive $\eta$ is close to zero   the outer horizon radius becomes very small so that the ergosphere in the equatorial plane is much thicker than that in the case $a<M$, which could yield some new phenomena in the process of extracting energy from a Konoplya-Zhidenko rotating non-Kerr black hole.

\section{Energy extraction from a Konoplya-Zhidenko rotating non-Kerr black hole by Penrose process}

In this section, we will study  energy extraction from a Konoplya-Zhidenko rotating non-Kerr black hole by Penrose process and to see the effect of
the deformation parameter $\eta$ on the negative energy state and on the energy extraction efficiency .

For simplicity, we focus only on the orbit of a test particle with the mass $\mu$ on the equatorial plane. Making use of the two Killing vectors $\xi^a=(\frac{\partial}{\partial t})$ and $\psi^a=(\frac{\partial}{\partial \phi})$ of the spacetime (\ref{metric1}), one can obtain two conserved quantities for the particles moving along a timelike geodesics
\begin{eqnarray}
E&=&-g_{ab}\xi^au^b=\bigg(1-\frac{2Mr^2+\eta}{ r^3}\bigg)u^t+\frac{2Mr^2+\eta}{r^3}a u^\phi, \\
L&=&g_{ab}\psi^au^b=-\frac{2Mr^2+\eta}{r^3}a u^t+\bigg[(r^2+a^2)+\frac{(2Mr^2+\eta)a^2}{r^3}\bigg]u^\phi.
\end{eqnarray}
Here $u^b\equiv\frac{dx^b}{d\tau}$ is the four-velocity and $\tau$ is the proper time for the spacetime. The conserved quantities $E$ and $L$ correspond to the energy or angular momentum of the particle  \cite{L10,L11,L12}, respectively.
For a timelike particle in the background of a Konoplya-Zhidenko rotating non-Kerr black hole, one can find the energy $E$ of the particle moving along geodesics satisfies
\begin{eqnarray}\label{13}
\alpha E^2-2\beta E+\gamma&=&0,
\end{eqnarray}
with
\begin{eqnarray}\label{14a}
\alpha&=&(r^2+a^2+\frac{2Ma^2}{r}+\frac{\eta a^2}{r^3})\Gamma^{-1}, \\
\beta&=&L(\frac{2Mr^2+\eta}{r^3}a)\Gamma^{-1}, \label{14b}\\
\gamma&=&-L^2(1-\frac{2Mr^2+\eta}{r^3})\Gamma^{-1}-\frac{r^3}{r^3-2Mr^2+a^2r -\eta}(u^r)^2-\mu^2,\label{14c}\\
\Gamma&=&\frac{r^3-2Mr^2+a^2r-\eta}{r}.\label{14d}
\end{eqnarray}
Solving Eq.(\ref{13}), one can find the energy $E$ has a form
\begin{eqnarray}\label{18}
E&=&\frac{\beta\pm\sqrt{\beta^2-\alpha\gamma}}{\alpha}.
\end{eqnarray}
In order to ensure that the 4-momentum of the particle is future directed, we select only the sign $(+)$ in front of $ \sqrt{\beta^2-\alpha\gamma}$ for the energy $E$ in the following discussion. It is well known that the orbit of the particle with negative energy in the ergosphere is crucial to extract energy from a rotation black hole through Penrose process.
From Eq.(\ref{18}), one can find that the conditions of the particle with the negative energy (i.e., $E<0$) are: $\alpha>0$, $\beta<0$ and $\gamma>0$, which can be satisfied only if $La<0$. As a negative energy particle with mass $\mu$ is injected into the central black hole, the mass of the black hole will change a quantity $\delta M=E$. It is shown that there exists a lower limit on $\delta M$ which could be added to the black hole corresponding to the case with $\mu=0$ and $\mu^r=0$ \cite{L15}. The lower limit $E_{\min}$ can be evaluated through all of the required quantities at the horizon $r_H$,
\begin{eqnarray}\label{19}
E_{\min}&=&\frac{L(2Mr_H^2+\eta)a}{(r_H^2+a^2)r^3_H+(2Mr_H^2+\eta)a^2}
=\frac{La}{r^2_H+a^2}.
\end{eqnarray}
This implies that the energy can be extracted from the black hole by Penrose process only if the injected particle possesses the negative angular momentum (i.e., $L<0$). Moreover, one can find the value of $E_{min}$ depends on the deformation parameter $\eta$ since the horizon radius $r_H$ is a function of $\eta$.
\begin{figure}[ht]
\includegraphics[width=5.2cm]{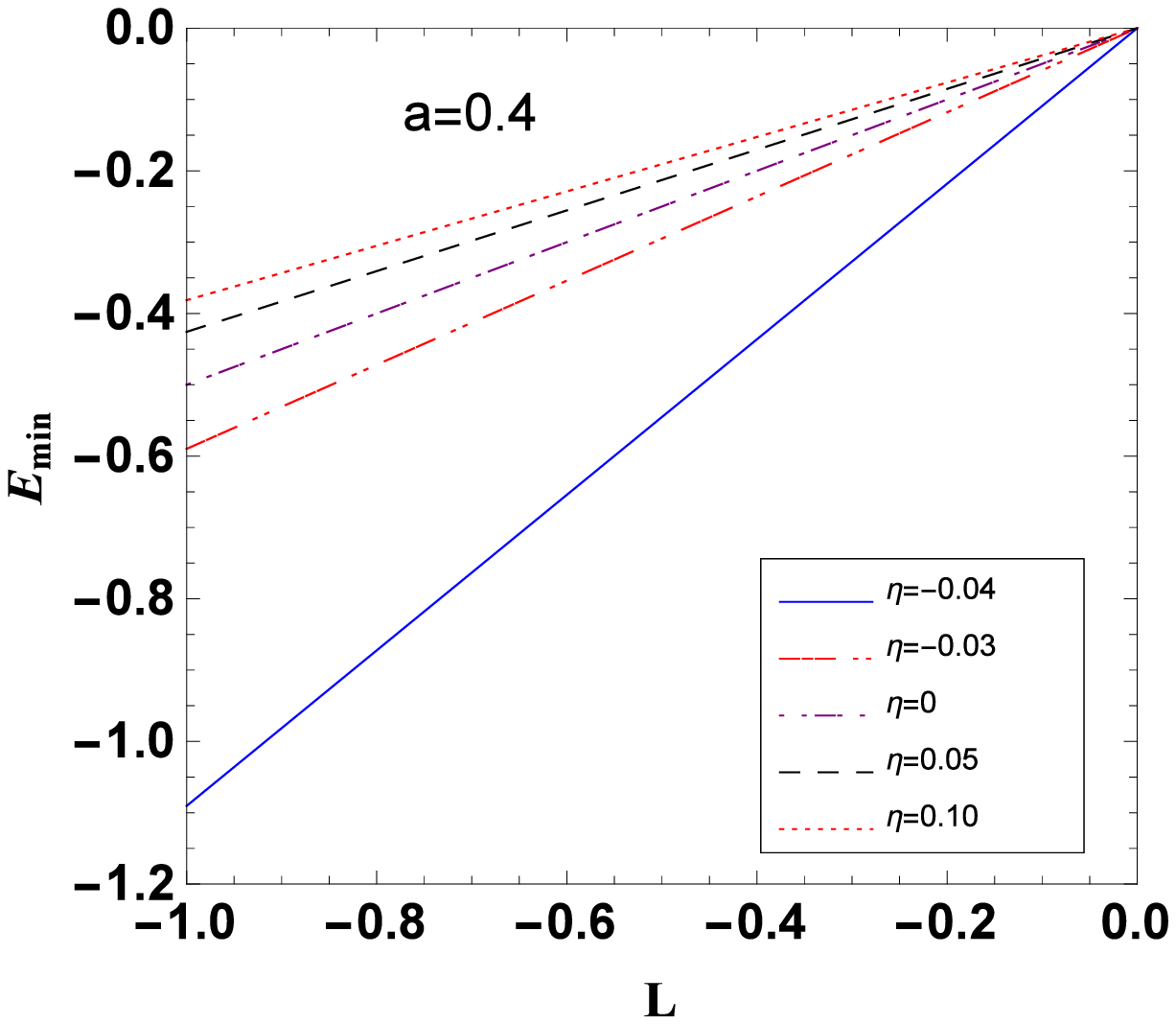}
\;\;\;\;\includegraphics[width=5.2cm]{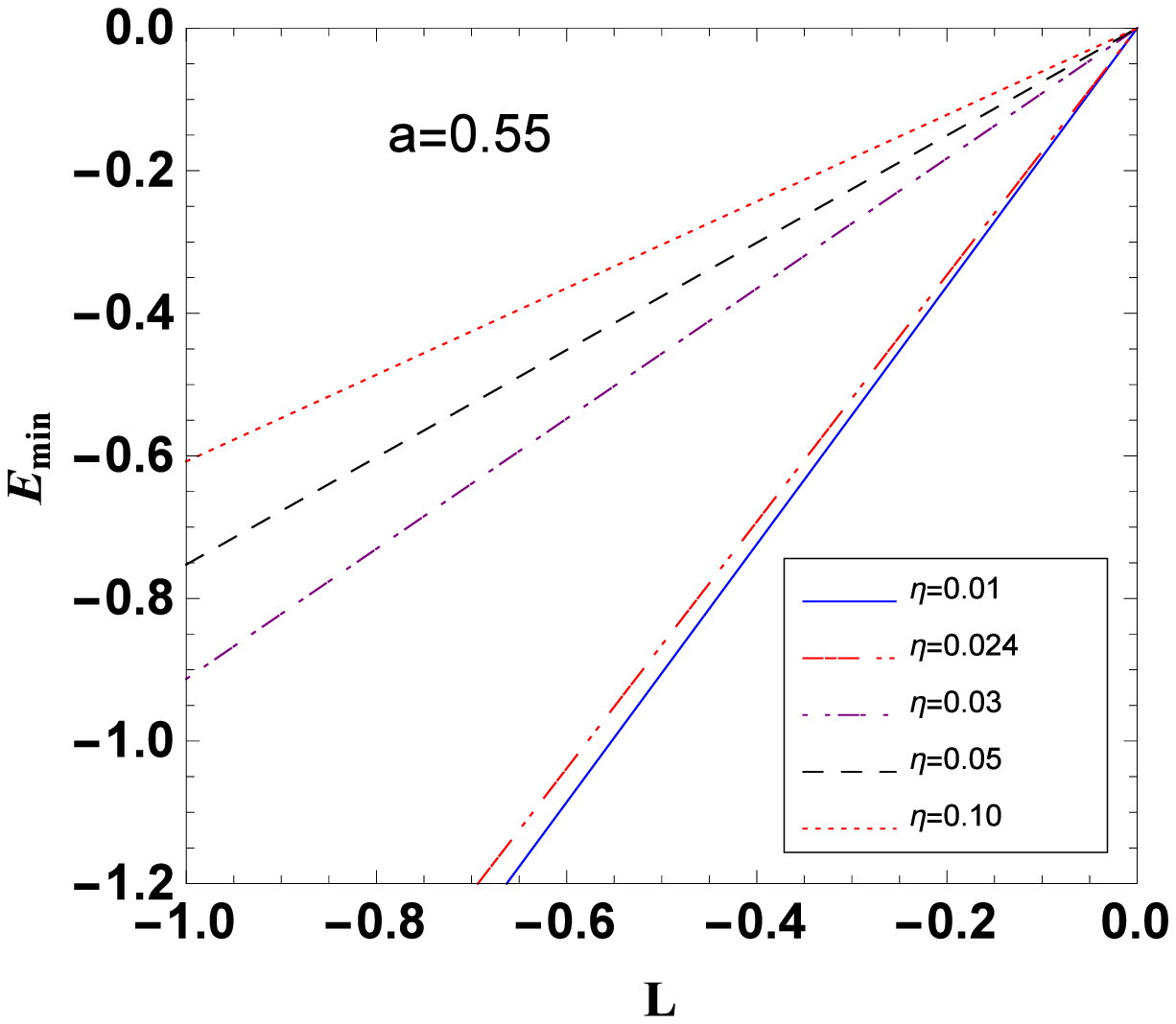}
\;\;\;\;\includegraphics[width=5.2cm]{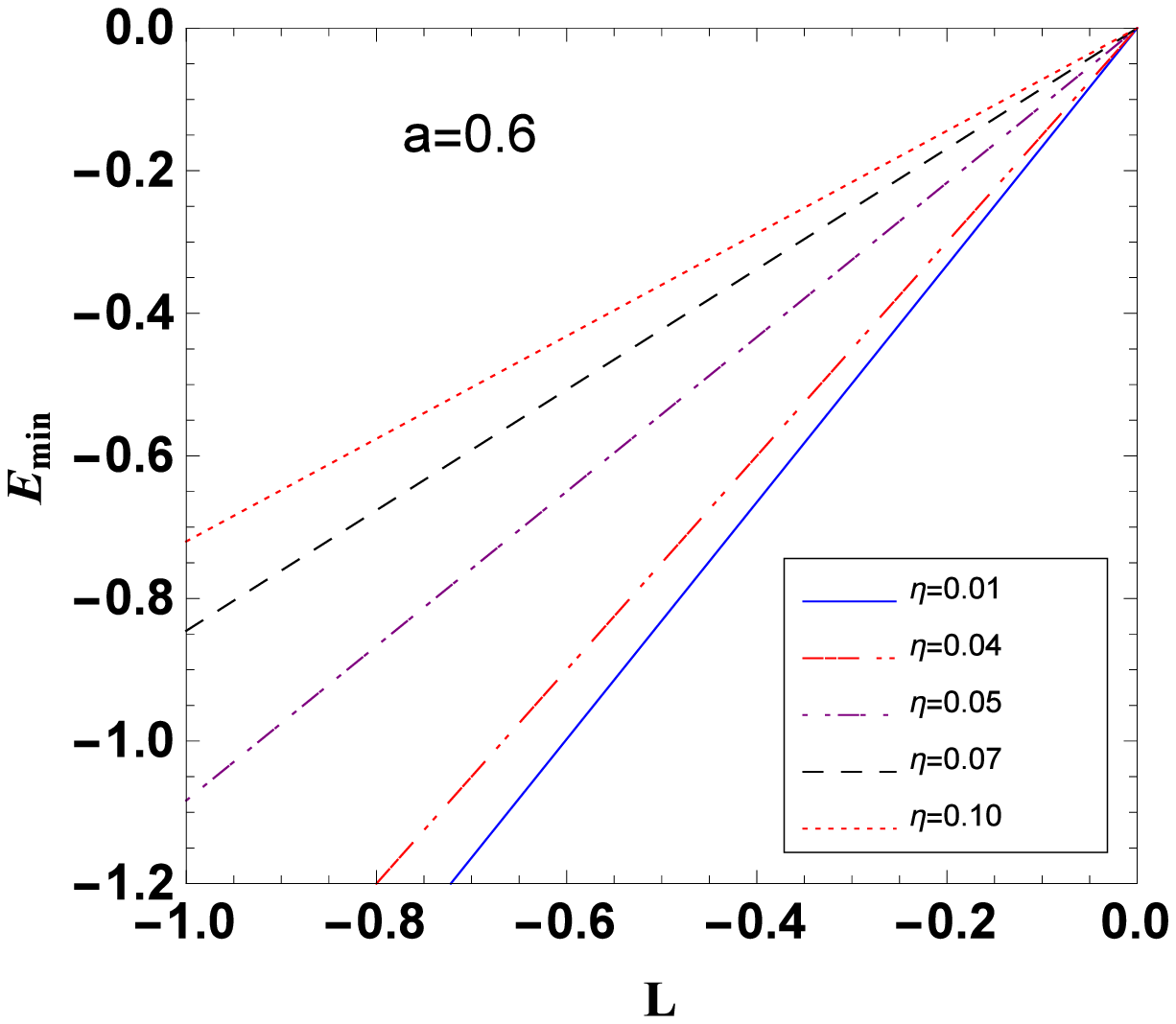}
\caption{The change of the lower limit $E_{\min}$ of negative energy state with the angular momentum $L$ for the different  deformation parameter $\eta$ and rotation parameter $a$ in the Konoplya-Zhidenko rotating non-Kerr black hole spacetime.}
\end{figure}
In Fig.(5), we present the change of the lower limit $E_{\min}$ with the angular momentum $L$ for the different  deformation parameter $\eta$ and rotation parameter $a$ in the Konoplya-Zhidenko rotating non-Kerr black hole spacetime. It's easy to see that the absolute value of $E_{\min}$ decrease monotonously with the deformation parameter $\eta$ for all $a$.

Let us now to study the effect of the deformation parameter $\eta$ on the efficiency of the energy extraction process in the Konoplya-Zhidenko rotating non-Kerr black hole spacetime. In order to calculate the maximum efficiency of the energy extraction in this spacetime, we take the radial velocity to be zero as in Refs.\cite{L10,L11,L12}. In the locally nonrotating frame \cite{W1}, the four-velocity $U_i$ of the $i$th particle for the observer at a given radius $r$ can be expressed as
\begin{eqnarray}
U_i=u^t(1,0,0,\Omega_i),
\end{eqnarray}
with
\begin{eqnarray}
u^t=\frac{E}{g_{tt}+g_{t\phi}\Omega_i},\;\;\;\;\;\;\;\;\;\;\;\;
\Omega_i=\frac{-g_{t\phi}(1+g_{tt})+\sqrt{(1+g_{tt})(g^2_{t\phi}
-g_{tt}g_{\phi\phi})}}{g_{\phi\phi}+g^2_{t\phi}},
\end{eqnarray}
where $\Omega_i$ is the angular velocity of the  $i$th  particle with respect to an asymptotic infinity observer. In the ergosphere, the value of $\Omega_i$ lies in the range of
\begin{eqnarray}
\Omega_-<\Omega_i<\Omega_+,
\end{eqnarray}
where
\begin{eqnarray}
\Omega_{\pm}=\frac{-g_{t\phi}\pm\sqrt{g^2_{t\phi}-
g_{tt}g_{\phi\phi}}}{g_{\phi\phi}}.
\end{eqnarray}
 According to the Penrose process \cite{L10,L11,L12}, an incident particle $1$ with the rest mass $\mu_1=1$ (i.e., $E_1=1$)  after entering the ergosphere splits into two particles $2$ and $3$, and then the consequence will be that the particle $2$ with negative energy falls past the outer event horizon into the black hole, while the particle 3 escapes to infinity with  more energy than the incident particle $1$.
 From the conservational laws of the energy and angular momentum, one can obtain
\begin{eqnarray}
U_1=\mu_2U_2+\mu_3U_3.
\end{eqnarray}
And then the efficiency of the energy extraction in Penrose process can be expressed as
\begin{eqnarray}
\epsilon=\frac{\mu_3E_3-E_1}{E_1}=\mu_3E_3-1.
\end{eqnarray}
As in Ref.\cite{L12,chen2,chen201}, choosing $\mu_2U_2$ and $\mu_3U_3$ as the forms
\begin{eqnarray}
\mu_2U_2=k_2(1,0,0,\Omega_-),\nonumber\\
\mu_3U_3=k_3(1,0,0,\Omega_+),
\end{eqnarray}
with two undetermined constants $k_2$ and $k_3$,
one can obtain the efficiency $\epsilon$ of energy extraction from a rotation black hole
\begin{eqnarray}
\epsilon=\frac{(\Omega_1-\Omega_-)(g_{tt}+g_{t\phi}\Omega_+)}{
(\Omega_+-\Omega_-)(g_{tt}+g_{t\phi}\Omega_1)}-1.
\end{eqnarray}
The maximum efficiency can be evaluated by assuming that the incident particle 1 splits near the horizon $r_H$,
\begin{eqnarray}\label{xlv}
\epsilon_{\max}=\frac{\sqrt{1+g_{tt}}-1}{2}|_{r=r_H}.
\end{eqnarray}
\begin{figure}[ht]
\begin{center}
\includegraphics[width=5.3cm]{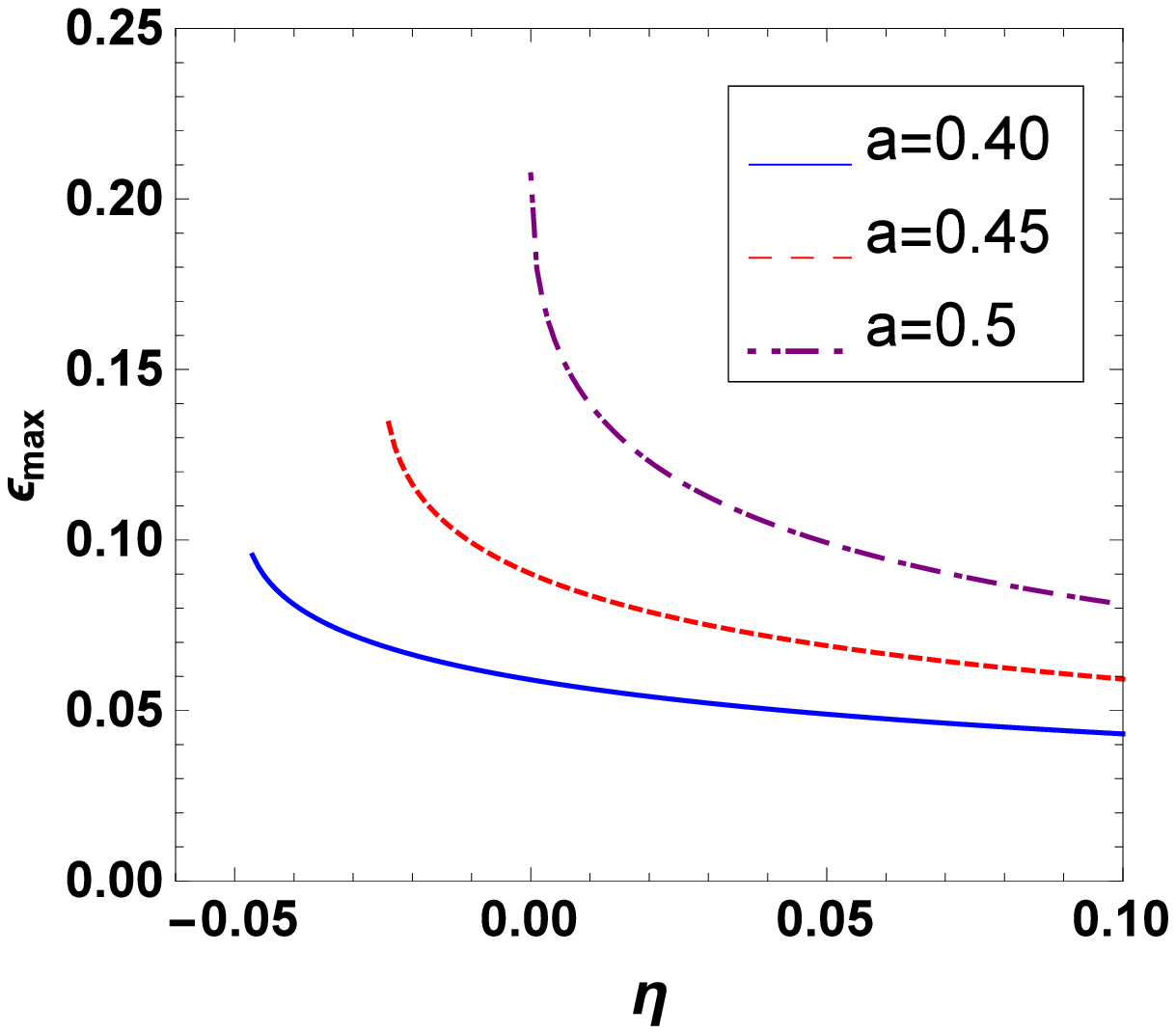}
\;\;\;\;\includegraphics[width=5.2cm]{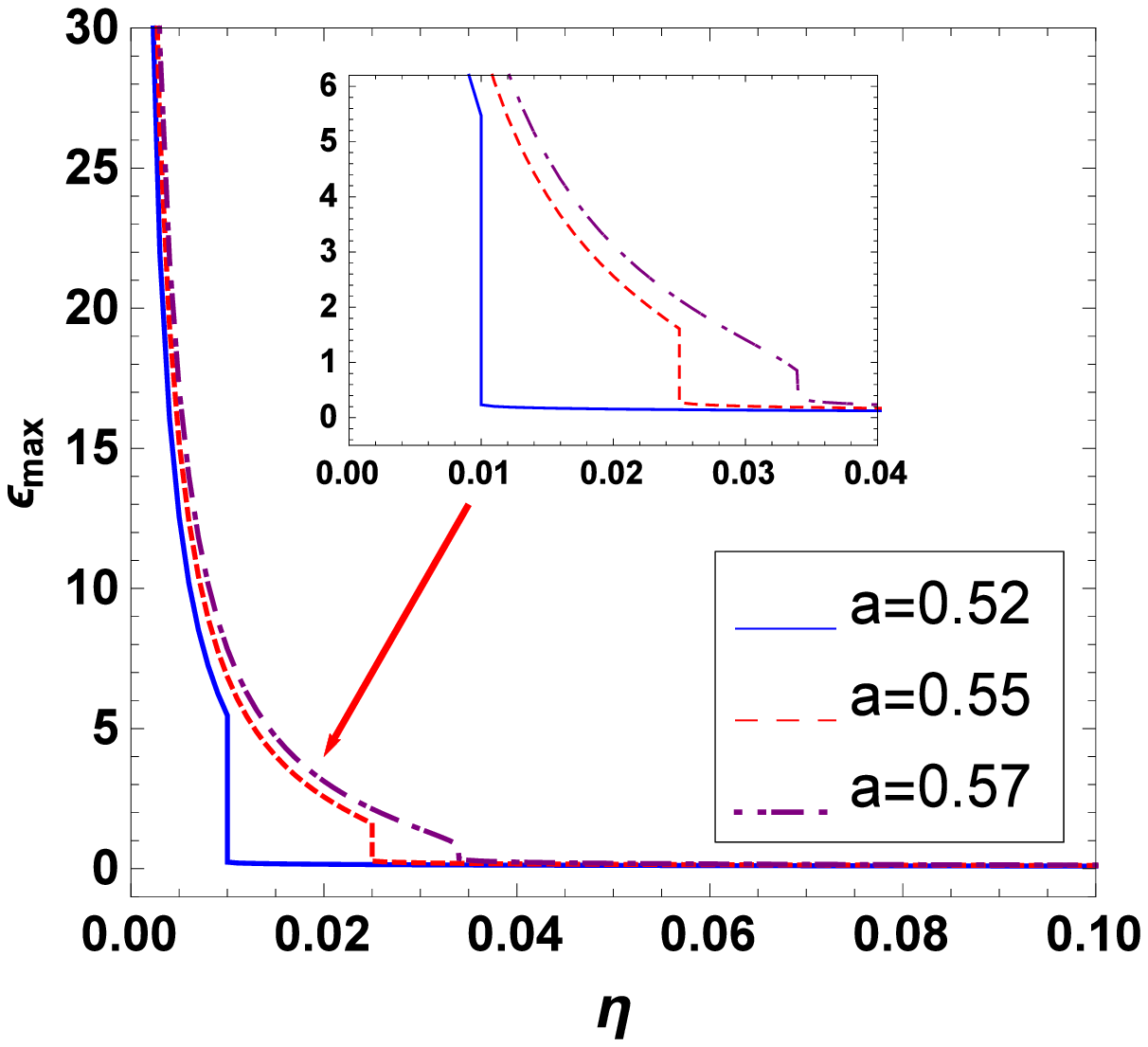}
\includegraphics[width=5.2cm]{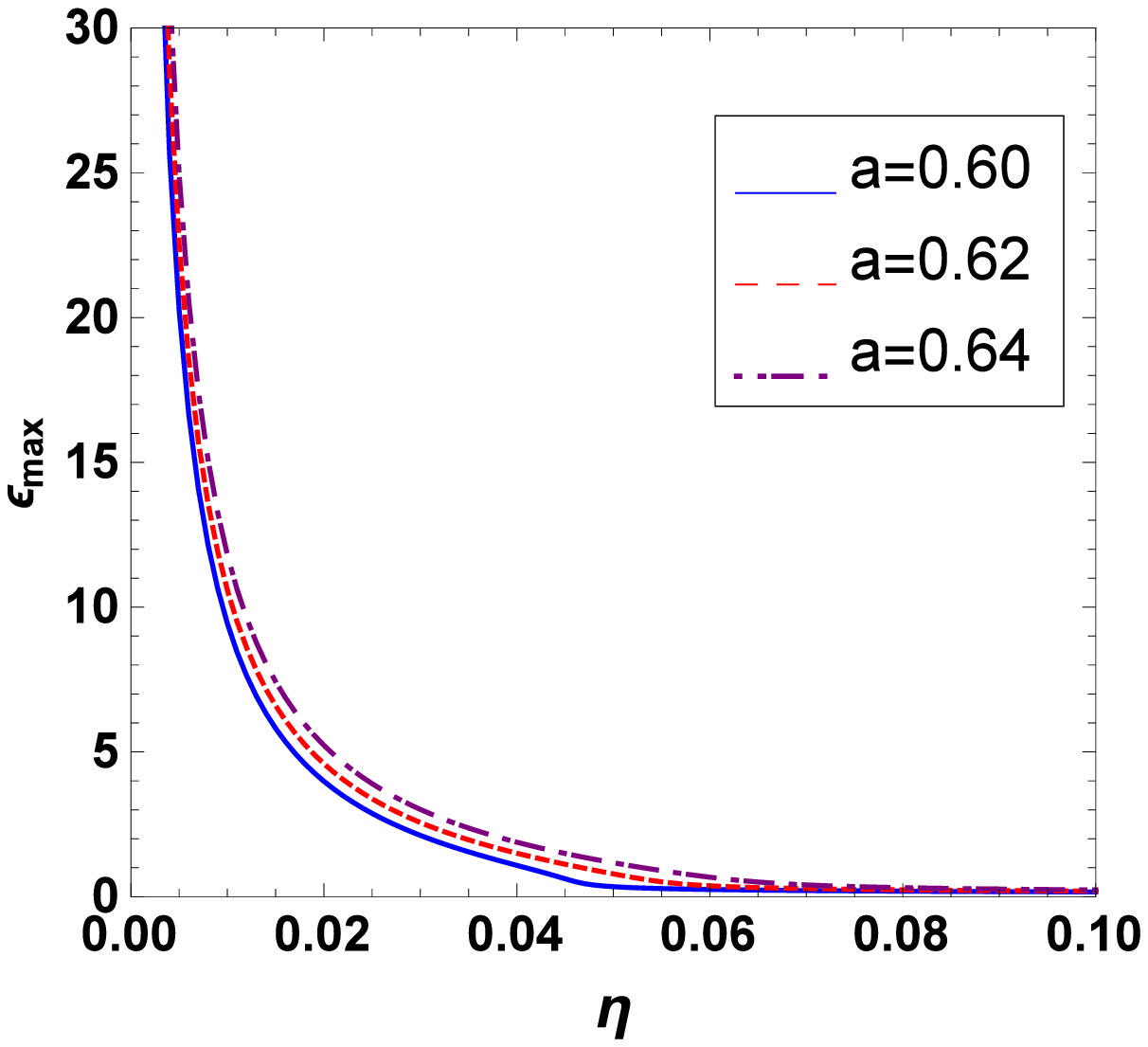}
\caption{The variation of the maximum efficiency of the energy extraction with the deformation parameter $\eta$ from  a Konoplya-Zhidenko rotating non-Kerr black hole. Here we take $M=0.5$. }
\end{center}
\end{figure}
Substituting the coefficients of the  metric (\ref{metric1}) into Eq.(\ref{xlv}), we can analyze the effects of $\eta$ on the energy extraction efficiency from a Konoplya-Zhidenko rotating non-Kerr black hole. With the help of the numerical method, we present the variation of the maximum efficiency in the energy extraction process with the deformation parameter $\eta$, which takes the values so that there exist event horizon for the non-Kerr metric (\ref{metric1}). From Fig.(6), we find that the maximum efficiency of the Penrose process decreases with the increase of the deformation parameter $\eta$ in the Konoplya-Zhidenko rotating non-Kerr spacetime, which differs  from that in the Johannsen-Psaltis rotating non-Kerr spacetime in which the maximum efficiency is an increasing function of the deformation parameter \cite{chen2,chen201}. For the case with $a<M$, the positive deformation parameter yields that the maximum efficiency is lower than that in the Kerr black hole with the same rotation parameter,  while the negative deformation parameter enhances the maximum efficiency of the energy extraction process greatly. For the superspinning case with $a>M$, it
is of interest to note that the maximum efficiency can reach so high that it is almost unlimited as the positive $\eta$ is close to zero, while in the Einstein's theory of gravity the maximum efficiency is only $20.7\%$ for the extremal Kerr black hole.
 The almost unlimited efficiency is could be explained by a fact that in this case the outer event horizon radius is very small so that the maximum efficiency (\ref{xlv}) can be approximated as $\epsilon_{\max}\approx \sqrt{\frac{2M}{r_H}}$. Moreover, the ergosphere in the equatorial plane becomes much thicker in this special case. Actually, the almost unlimited efficiency is also found in the energy extraction process from a Johannsen-Psaltis rotating non-Kerr black hole as the negative $\eta$ is close to zero in the case with $a>M$ \cite{chen2,chen201}. Thus,
the emergence of the almost unlimited efficiency may be a new feature of energy extraction in such kind of rotating non-Kerr black hole spacetime with $a>M$. From the middle panel in Fig.(6),  we find that there exists a sudden change for the maximum efficiency in the cases with $M<a<\frac{2\sqrt{3}M}{3}$ as the value $\eta$ crosses the critical value $\eta_2$ since the variation of the event horizon $r_H$ is not continuous (which is also shown in Fig.(1)). Actually, the similar sudden change also takes place for the width of the ergosphere as in the previous discussion.

\section{Summary}
In this paper we have investigated the energy extraction from a Konoplya-Zhidenko rotating non-Kerr black hole with an extra deformation parameter. We find that the deformed
parameter together with the rotation parameter imprint in the properties of ergosphere and the energy extraction efficiency. With the increase of the deformation parameter, the ergosphere becomes thin and the maximum efficiency of energy extraction decreases, which differs  from that in the Johannsen-Psaltis rotating non-Kerr spacetime where the maximum efficiency increases with the deformation parameter. For the case with $a<M$, the positive deformation parameter yields that the maximum efficiency is lower than that in the Kerr black hole with the same rotation parameter,  while the negative deformation parameter enhances the maximum efficiency of the energy extraction process greatly. For the superspinning case with $a>M$, we find that the maximum efficiency can reach so high that it is almost unlimited as the positive $\eta$ is close to zero, which is different from that in the usual Kerr black hole spacetime in which the maximum efficiency is only $20.7\%$ for the extremal Kerr black hole.
Our result imply that the emergence of the almost unlimited efficiency may be a new feature of energy extraction in such kind of rotating non-Kerr black hole spacetime with $a>M$. Finally, we find that in the cases with $M<a<\frac{2\sqrt{3}M}{3}$  there exists a sudden change for the maximum efficiency and the width of the ergosphere as the value $\eta$ crosses the critical value $\eta_2$ since the variation of the event horizon $r_H$ is not continuous.
These effects of the deformation parameter $\eta$ on the maximum efficiency
could provide a possibility to check the no-hair theorem and to test whether or not the current black-hole candidates are the Konoplya-Zhidenko rotating non-Kerr black holes beyond Einstein's General Relativity in the future astronomical observations.

\section{\bf Acknowledgments}

This work was partially supported by the Scientific Research
Fund of Hunan Provincial Education Department Grant
No. 17A124. J. J.'s work was partially supported by
the National Natural Science Foundation of China under
Grant No. 11475061.


\vspace*{0.2cm}
\begin{thebibliography}{99}

\baselineskip=0.6 cm \baselineskip=0.6 cm

\bibitem{W2} B.  P.  Abbott \textit{et al}. ,  (LIGO Scientific and Virgo Collaborations),    Phys.  Rev.  Lett.  {\bf116}, 061102 (2016),
arXiv:1602. 03837 [gr-qc].
\bibitem{W21}B. P. Abbott  \textit{et al}. (LIGO Scientific Collaboration, Virgo Collaboration), Phys. Rev. Lett.  {\bf116} 116, 241103 (2016),
arXiv:1606.04855 [gr-qc].
\bibitem{W31} B.  P.  Abbott et al. ,  [The LIGO Scientific and the Virgo Collaborations], Phys. Rev. Lett. {\bf118} (2017) 221101,  arXiv:1706.01812 [gr-qc].


\bibitem{L1} W. Israel, Phys. Rev. {\bf164}, 1776 (1967); W. Israel, Commun. Math. Phys. {\bf8}, 245 (1968); B. Carter, Phys. Rev. Lett. {\bf26},  331(1971); S. W. Hawking, Commun. Math. Phys. {\bf25}, 152 (1972); D. C. Robinson, Phys. Rev. Lett. {\bf34}, 905 (1975).
\bibitem{W1} C.  M.  Will, Living Rev. Rel. {\bf9}, 3 (2005);  Living Rev.  Rel.  {\bf17}, 4  (2014).
\bibitem{W3} B.  P.  Abbott et al. ,  [The LIGO Scientific and the Virgo Collaborations], Phys.  Rev.  Lett.  {\bf116} (2016) 221101,  arXiv:1602.03841 [gr-qc].

\bibitem{fR} A. De Felice and S. Tsujikawa,  Living Rev. Rel. {\bf13}, 3 (2010).

\bibitem{sd1} Supernova Search Team collaboration, A. G. Riess \textit{et al}.,  Astron. J. {\bf116}, 1009 (1998).
\bibitem{sd2} A. G. Riess \textit{et al}.,  Astrophys. J. {\bf659}, 98 (2007).
\bibitem{sd3} Boomerang collaboration, P. de Bernardis \textit{et al}.,  Nature {\bf404}, 955 (2000).
\bibitem{sd4}Supernova Cosmology Project collaboration, S. Perlmutter \textit{et al}., Astrophys. J. {\bf517}, 565 (1999).
\bibitem{sd5}Supernova Cosmology Project collaboration, R. A. Knop \textit{et al}., Astrophys. J. {\bf598}, 102 (2003).

\bibitem{JP1} T.Johannsen and D. Psaltis, Phys. Rev. D {\bf83}, 124015 (2011).
\bibitem{CBa1} C. Bambi and L. Modesto, Phys.Lett.B {\bf706},13 (2011).

\bibitem{Cos101} C. Bambi, Phys. Rev.D {\bf85},  043001 (2012).
 \bibitem{Cos102}C. Bambi, Astrophys. J. {\bf761}, 174 (2012).

\bibitem{chen1} S. Chen and J. Jing, Phys. Lett. B {\bf711}, 81 (2012).

 \bibitem{chen12} S. Chen and J. Jing, Phys. Rev. D {\bf85}, 124029  (2012).


\bibitem{Kraw} H. Krawczynski, Astrophys. J. {\bf754}, 133 (2012).

\bibitem{Ra1}  R. A. Konoplya and A. Zhidenko,  Phys. Rev. D {\bf87},  024044 (2013).

 \bibitem{Fa1} F. Atamurotov, A. Abdujabbarov and B. Ahmedov,  Phys. Rev. D{\bf88}, 064004 (2013).


\bibitem{And1}A. Maselli, K. Kokkotas and P. Laguna, Phys. Rev. D {\bf93}, 064075 (2016).


\bibitem{Test3}C. Bambi,  JCAP {\bf1209}, 014  (2012); Eur. Phys. J. C {\bf75}  162 (2015);
\bibitem{Test302}L. Kong, Z. Li and C. Bambi, Astrophys. J. {\bf797}, 78 (2014).
\bibitem{Test303} J. Jiang, C. Bambi and J. F. Steiner, JCAP {\bf1505}, 025 (2015).

\bibitem{Test304} D. Liu, Z. Li and C. Bambi, JCAP {\bf1501}, 020 (2015).
\bibitem{Test305} C. Bambi, JCAP {\bf1308}, 055 (2013).

\bibitem{Test1} T. Johannsen and D. Psaltis, Astrophys. J. {\bf773}, 57 (2013). 

\bibitem{Test2} C. Bambi,  Phys. Lett. B {\bf705}, 5 (2011).
\bibitem{Test201} C. Bambi, Phys. Rev. D {\bf85}, 043002 (2012).
\bibitem{Test202} C. Bambi, Phys. Rev. D {\bf87}, 023007 (2013).

\bibitem{W4} R.  Konoplya and A.  Zhidenko,  Phys.  Lett.  B {\bf756}, 350 (2016).

\bibitem{GKt01} C. Bambi and S. Nampalliwar, Europhys. Lett. {\bf116}, 30006 2016.
\bibitem{GKt02} Y. Ni, J. Jiang and C. Bambi, JCAP {\bf09},  014 (2016).
\bibitem{W6} S. Wang, S. Chen and J. Jing, J. Cosmol. Astropart. Phys. {\bf11}, 020 (2016).

\bibitem{L10} R. Penrose, Riv. Nuovo Cumero Speciale {\bf1}, 252  (1969).

\bibitem{L11} S. Chandrasekhar, \textit{The Mathematical Theory of black holes} (Oxford University Press, New York, 1983).

\bibitem{L12} M. Bhat, S. Dhurandrhar and N. Dadhich, Astrophys. J. {\bf6}, 85 (1985); S. Parthasarathy, S. M. Wagh, S. V. Dhurandhar and N. Dadhich, Astrophys. J. {\bf307}, 38 (1986).
    

\bibitem{chen2} C. Liu, S. Chen and J. Jing, Astrophys. J. {\bf751}, 148 (2012).
\bibitem{chen201} S. G. Ghosh and P. Sheoran,  Phys. Rev. D {\bf89}, 024023 (2014).

\bibitem{L15} C. W. Wisner, K.S. Thore, and J.A. Wheeler, \textit{Gravitation} (SanFrancisco, Freeman, 1973).


\end{thebibliography}
\end{document}